\pdfoutput=1
\documentclass{article}
\usepackage{arxiv}
\usepackage[T1]{fontenc}
\usepackage[utf8]{inputenc}
\usepackage{float}
\usepackage{graphicx}
\usepackage{subcaption}
\usepackage{booktabs}
\usepackage{array}
\usepackage{tabularx}
\usepackage{amsmath}
\usepackage{xcolor}
\usepackage{amssymb}
\usepackage[hidelinks]{hyperref}
\usepackage{adjustbox}
\usepackage{etoolbox}

\newcommand{\tablefont}{\small}
\newcommand{\tablewidth}{0.9\linewidth}

\AtBeginEnvironment{table}{%
    \centering
    \tablefont
    \setlength{\tabcolsep}{6pt}%
    \renewcommand{\arraystretch}{1.15}%
}

\newenvironment{alignedpanel}{%
  \begin{minipage}[c][0.85\linewidth][c]{\linewidth}\centering
}{\end{minipage}\par}

\newenvironment{pairedfieldpanel}{%
  \begin{minipage}[c][1.45\linewidth][c]{\linewidth}\centering
}{\end{minipage}\par}
\newcommand{\pairedfieldviews}[1]{%
  \includegraphics[width=\linewidth,height=0.70\linewidth,keepaspectratio]{#1.pdf}\par\vspace{0.025\linewidth}%
  \includegraphics[width=\linewidth,height=0.70\linewidth,keepaspectratio]{#1_stress.pdf}%
}

\title{Mixed-Dimensional Electromechanical Coupling of Embedded Beam Networks in Soft Dielectric Composites}
\renewcommand{\shorttitle}{Mixed-Dimensional Electromechanical Beam Networks}

\author{
L. River Spencer \\
Department of Aerospace Engineering \& Engineering Mechanics,\\
The University of Texas at Austin,\\
Austin, TX 78712 \\
\And
Manuel K. Rausch \\
Department of Aerospace Engineering \& Engineering Mechanics,\\
Department of Biomedical Engineering, \\
The Oden Institute of Computational Science and Engineering,\\
The University of Texas at Austin,\\
Austin, TX 78712 \\
\And
Chad Landis \\
Department of Aerospace Engineering \& Engineering Mechanics,\\
The Oden Institute of Computational Science and Engineering,\\
The University of Texas at Austin,\\
Austin, TX 78712 \\
\And
Jan N. Fuhg \\
Department of Aerospace Engineering \& Engineering Mechanics,\\
The Oden Institute of Computational Science and Engineering,\\
The University of Texas at Austin,\\
Austin, TX 78712
}

\date{}
\begin{document}

\maketitle

\begin{abstract}
Soft electroactive composites can exhibit strongly architecture-dependent mechanical and electrical responses, but explicitly resolving dense fiber networks in three dimensions is computationally expensive. We develop a mixed-dimensional finite-element formulation in which electroactive fibers are represented by geometrically exact beams embedded in a deformable dielectric matrix. Unlike existing electroactive beam formulations in which the electric potential is defined directly on the beam, the embedded fibers here are driven by the three-dimensional electric field of the surrounding matrix. The matrix field is sampled along the beam centerlines and enters the beam dielectric enthalpy directly, providing two-way electromechanical coupling without introducing independent electric-potential degrees of freedom on the beams. Beam and matrix mechanics are coupled through projected mortar constraints, and the electromechanical problem is solved monolithically. We use the formulation as the microscale model in a periodic homogenization framework to compute effective stress and electric displacement. Verification studies quantify discretization and field-sampling sensitivity, followed by structured and irregular network examples that demonstrate architecture-dependent mechanical and electrical responses without body-fitted three-dimensional fiber meshes.
\end{abstract}

\section{Introduction}\label{sec:intro}
Dielectric elastomers combine large mechanical deformation with an electrical
response and provide a basis for soft actuation and deformation sensing
\cite{pelrine2000high,ohalloran2008review}. Mechanical deformation can also
change electrical quantities such as capacitance, enabling sensing and
self-sensing applications
\cite{jung2008self,bose2023sensors,rizzello2023review}. In heterogeneous
dielectric elastomers, these coupled responses depend on the constituent
properties as well as on the local mechanical and electric fields
\cite{kanan2021electroviscoelastic,kanan2021heterogeneous}. Fibrous phases
provide an additional means of tailoring this response through their
architecture. Passive fiber reinforcement has been used to control actuation
direction, force, and deformation mode
\cite{huang2012large,moss2021modeling,holzer2024fiber}, while fibers can
also form an electroactive phase themselves \cite{kang2024fiber}. In both
cases, fiber orientation, curvature, connectivity, and density can influence
the interaction between the network and the surrounding matrix and therefore
the effective response of the composite. These architectural features can also be controlled through processing. Electrospinning, for example, provides a
route for generating fibrous networks with controlled orientation and
alignment \cite{greiner2007electrospinning,xue2019electrospinning}.
Oriented electrospun fiber composites have already been used to produce
anisotropic dielectric-elastomer actuation
\cite{zhang2023anisotropicElectrospunDEA}, while collection strategies can
be used to generate aligned fiber architectures
\cite{li2003alignedElectrospun}. This creates the possibility of treating network architecture as a
material-design variable. Exploring that design space requires computational
models that retain the relevant network features. A direct approach is to
represent both the fibers and the surrounding matrix with three-dimensional
continuum elements. Such a body-fitted discretization must resolve slender
fiber volumes, their interfaces with the matrix, and potentially large
numbers of junctions. The computational cost therefore increases rapidly
for dense networks with small fiber radii and large aspect ratios. The modeling challenge is therefore to retain the explicit architecture of an electroactive fiber network without resolving the full three-dimensional fiber geometry, while still accounting for its coupled mechanical and electrical interaction with the surrounding matrix.

\begin{figure}[!htbp]
    \centering
    \includegraphics[width=0.8\textwidth]
    {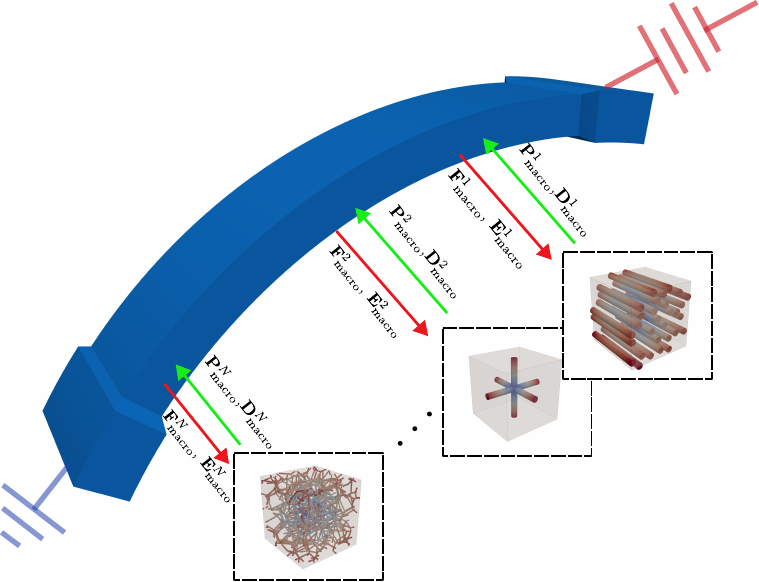}
    \caption{Mixed-dimensional periodic homogenization of embedded electroactive fiber networks. One-dimensional dielectric beams are mechanically coupled to a three-dimensional matrix, while their electrical response is driven by the surrounding matrix field. The resulting periodic RVE problem provides homogenized stress and electric displacement.}
    \label{fig:graphical-abstract}
\end{figure}
Mixed-dimensional formulations provide a natural starting point for this
problem by representing slender inclusions through lower-dimensional
structural kinematics coupled to a three-dimensional solid
\cite{mccune2000mixed,sadek2004embedded}. Beam--solid formulations have
developed pointwise and mortar-based strategies for coupling nonmatching
beam and solid discretizations
\cite{ninic2014beam,steinbrecher2020mortar}, while variational approaches
have been proposed for fiber-reinforced solids
\cite{khristenko2021multidimensional}. Embedded beam networks have also
been used to investigate how discrete fiber architecture and the surrounding
matrix jointly determine the mechanical response
\cite{kakaletsis2023mechanics}. These developments provide the mechanical
basis for retaining an explicit network without constructing a
three-dimensional mesh of every fiber.
For electroactive fibers, however, reducing the mechanical geometry is only part of the problem: the electric field acting on the reduced structure must also be defined. Existing electromechanically coupled beam models introduce electric-potential degrees of freedom directly on the beam \cite{HuangLeyendecker2022}, consistent with beams that are individually addressed by electrodes. The embedded fibers considered here are instead driven by the three-dimensional electric field generated in the surrounding matrix. We therefore evaluate the matrix electric field along the beam centerlines and use it directly in the beam dielectric enthalpy. Its dependence on both the matrix potential and beam deformation produces two-way electromechanical coupling without introducing an independent beam potential. To the best of our knowledge, this type of coupling between an embedded electroactive beam network and a surrounding three-dimensional dielectric continuum has not previously been reported.

The resulting mixed-dimensional formulation is not restricted to a
particular spatial scale. In the present work, it is used as the microscale
model within a periodic multiscale framework. Computational homogenization
provides the connection between heterogeneous microscopic fields and
effective material response
\cite{kouznetsova2001micro,geers2010multiscale,saeb2016aspects}, and this
micro-to-macro transition has been extended to finite-strain electroelastic
materials using coupled mechanical and electrical boundary conditions
\cite{schroder2012two,keip2014twoscale}. Related monolithic homogenization
strategies have also been developed for other coupled soft-material systems,
including nearly incompressible magnetoelastic composites
\cite{spencer2026monolithic}. Here, periodic displacement and
electric-potential fluctuations allow the local deformation and electric
fields to respond to the explicit fiber architecture. The equilibrated RVE
then provides homogenized stress and electric displacement obtained from the
complete coupled energy. Figure~\ref{fig:graphical-abstract} summarizes this
connection between the embedded network, the microscopic mechanical and
electrical fields, and the effective response available at the macroscale.
The numerical study begins with a single embedded fiber to assess
discretization sensitivity and to compare the additive reduced
representation with a body-fitted three-dimensional fiber. Structured and irregular networks are then used to examine
architecture-dependent mechanical and electrical responses and to
demonstrate the formulation for increasingly complex network geometries.

\section{Methods}\label{sec:methods}
The formulation couples a three-dimensional dielectric matrix to a nonmatching one-dimensional electroactive beam network. Mechanical compatibility is enforced through projected mortar coupling, while the beam dielectric response is driven by the matrix electric field sampled along the beam centerlines. We first present the general mixed-dimensional formulation and then specialize it to periodic homogenization in Sec.~\ref{sec:homogenization-background}.

\subsection{Problem Setting and Continuum Fields}
\label{sec:theoretical-setting}

%

Let $\Omega_0\subset\mathbb{R}^3$ denote the reference domain,
$\Omega_{m0}$ the matrix material domain, and $\Gamma_0$ the embedded beam
centerline network. For an unperforated domain, $\Omega_{m0}=\Omega_0$.
When a void is present, its reference domain is denoted by $\Omega_{v0}$,
with
$
\Omega_{m0}=\Omega_0\setminus\overline{\Omega}_{v0}.
$
The corresponding current domains are denoted by $\Omega$, $\Omega_m$, and
$\Omega_v$. The void is treated as an excluded insulating region: the matrix
displacement and electric-potential fields are defined only on $\Omega_{m0}$,
with traction-free and electrically insulating natural boundary conditions
on the internal matrix--void surface.

The matrix fields are
the displacement $\mathbf{u}_m(\mathbf{X})$ and electric potential
$\Phi(\mathbf{X})$. The beam fields are the centerline displacement
$\mathbf{u}_b(s)$ and rotation vector $\boldsymbol{\theta}_b(s)$, where
$s$ denotes the reference centerline coordinate. The embedded beams do not
carry an independent electric-potential field. Instead, their electrical
loading is obtained from the surrounding matrix electric field, as described
in Sec.~\ref{sec:matrix-projected-electric-coupling}. Unless stated
otherwise, vector components are expressed in the fixed orthonormal global
Cartesian basis $\{\mathbf e_1,\mathbf e_2,\mathbf e_3\}$.
The formulation is posed in a finite-deformation reference configuration
consistent with nonlinear field theories for deformable dielectrics
\cite{SuoZhaoGreene2008}. Let $\mathbf{X}$ and $\mathbf{x}$ denote the
positions of a material point in the reference and current configurations,
respectively. The motion $\boldsymbol{\varphi}$ and displacement field
$\mathbf{u}$ satisfy
\begin{equation}
\mathbf{x}(\mathbf{X})
=
\boldsymbol{\varphi}(\mathbf{X})
=
\mathbf{X}+\mathbf{u}(\mathbf{X}).
\label{eq:motion-definition}
\end{equation}
The associated deformation gradient, Jacobian, right Cauchy--Green tensor,
and first invariant are
\begin{equation}
\mathbf{F}
=
\nabla_{\!X}\boldsymbol{\varphi}
=
\mathbf{I}+\nabla_{\!X}\mathbf{u},
\qquad
J=\det\mathbf{F},
\qquad
\mathbf{C}=\mathbf{F}^T\mathbf{F},
\qquad
I_1=\mathrm{tr}\,\mathbf{C}.
\label{eq:finite-strain-kinematics}
\end{equation}
Here, $\mathbf I$ denotes the second-order identity tensor.
For the matrix, these quantities specialize to
\begin{equation}
\mathbf{F}_m(\mathbf{X})
=
\mathbf{I}+\nabla_{\!X}\mathbf{u}_m(\mathbf{X}),
\qquad
J_m=\det\mathbf{F}_m,
\qquad
\mathbf{C}_m=\mathbf{F}_m^T\mathbf{F}_m.
\label{eq:matrix-F-fluctuation}
\end{equation}
The electrostatic problem is written in terms of the total electric potential,
following the standard electroelastic description of dielectric solids
\cite{DorfmannOgden2005}. The referential electric field in the matrix is
\begin{equation}
\mathbf{E}_m(\mathbf{X})
=
-\nabla_{\!X}\Phi(\mathbf{X}).
\label{eq:electric-field-total-potential}
\end{equation}
The decomposition of $\mathbf{u}_m$ and $\Phi$ into prescribed macroscopic
parts and periodic fluctuations is introduced later when the formulation is
specialized to periodic computational homogenization in
Sec.~\ref{sec:boundary-load-control}.

\subsection{Matrix Electromechanical Model}
\label{sec:matrix-electromechanical-model}

The matrix is modeled as a nearly incompressible isotropic electroelastic
solid. To control volumetric deformation, its mechanical response combines
an isochoric neo-Hookean contribution with a volumetric penalty based on
element-wise J-bar averaging, following the energy split used in
\cite{spencer2026monolithic}. For a matrix element $\Omega_e$ with reference
volume $|\Omega_e|$, the element-averaged dilatation is
\begin{equation}
\bar{J}_e
=
\frac{1}{|\Omega_e|}
\int_{\Omega_e} J_m(\mathbf{X})\,dV.
\label{eq:jbar-cell-average}
\end{equation}
This averaged dilatation enters the volumetric energy, while the isochoric
response is evaluated from the pointwise volume-preserving part of the
matrix deformation gradient,
\begin{equation}
\widehat{\mathbf{F}}_m
=
J_m^{-1/3}\mathbf{F}_m,
\qquad
\det\widehat{\mathbf{F}}_m=1,
\qquad
\widehat{\mathbf{C}}_m
=
\widehat{\mathbf{F}}_m^T\widehat{\mathbf{F}}_m,
\qquad
\widehat{I}_{1m}
=
\mathrm{tr}\,\widehat{\mathbf{C}}_m.
\label{eq:jbar-isochoric-F}
\end{equation}
The resulting mechanical energy density and matrix contribution to the
potential are
\begin{equation}
\psi_m^{\mathrm{mech}}
=
\frac{K_m}{2}\left(\bar{J}_e-1\right)^2
+
C_{1m}\left(\widehat{I}_{1m}-3\right),
\qquad
\Pi_m^{\mathrm{mech}}
=
\int_{\Omega_{m0}}\psi_m^{\mathrm{mech}}\,dV,
\label{eq:matrix-mechanical-energy}
\end{equation}
where $K_m$ is the volumetric penalty coefficient and $C_{1m}$ is the
isochoric matrix coefficient. Because the volumetric energy depends on the
element average $\bar{J}_e$, this quantity must be differentiated together
with the deformation field,
\begin{equation}
\delta\bar{J}_e
=
\frac{1}{|\Omega_e|}
\int_{\Omega_e}
J_m\mathbf{F}_m^{-T}:\delta\mathbf{F}_m\,dV.
\label{eq:jbar-variation}
\end{equation}
Accordingly, the mechanical first Piola--Kirchhoff stress is defined through
\begin{equation}
\delta\Pi_{m,e}^{\mathrm{mech}}
=
\int_{\Omega_e}
\mathbf{P}_m^{\mathrm{mech}}:\delta\mathbf{F}_m\,dV,
\end{equation}
with
\begin{equation}
\mathbf{P}_m^{\mathrm{mech}}
=
K_m(\bar{J}_e-1)J_m\mathbf{F}_m^{-T}
+
2C_{1m}J_m^{-2/3}
\left(
\mathbf{F}_m
-
\frac{I_{1m}}{3}\mathbf{F}_m^{-T}
\right),
\qquad
I_{1m}=\mathrm{tr}(\mathbf{F}_m^T\mathbf{F}_m).
\label{eq:matrix-mechanical-piola}
\end{equation}
The J-bar quantity is therefore part of the variational response rather than
a frozen element-wise scalar. The present treatment is displacement based
and does not introduce a mixed pressure field; its discretization and
sensitivity to volumetric averaging are examined separately in the numerical
verification studies.

The same finite-deformation kinematics also determine the electrical
response of the matrix. We use the fixed-potential ideal-dielectric enthalpy
\begin{equation}
\psi_m^{\mathrm{elec}}
=
-\frac{1}{2}\epsilon_m J_m\,
\mathbf{E}_m\cdot\mathbf{C}_m^{-1}\mathbf{E}_m,
\qquad
\Pi_m^{\mathrm{elec}}
=
\int_{\Omega_{m0}}
\psi_m^{\mathrm{elec}}\,dV,
\label{eq:matrix-dielectric-energy}
\end{equation}
where $\epsilon_m$ is the matrix dielectric coefficient and
$\mathbf{E}_m$ is the referential electric field defined in
Sec.~\ref{sec:theoretical-setting}. Its negative derivative with respect to
the referential electric field gives
\begin{equation}
\mathbf{D}_m
=
-\frac{\partial\psi_m^{\mathrm{elec}}}{\partial\mathbf{E}_m}
=
\epsilon_m J_m\mathbf{C}_m^{-1}\mathbf{E}_m,
\label{eq:matrix-ideal-displacement}
\end{equation}
while differentiation with respect to deformation at fixed referential
electric field produces the electric contribution to the first
Piola--Kirchhoff stress. Introducing the spatial electric field
$\mathbf{e}_m=\mathbf{F}_m^{-T}\mathbf{E}_m$, this contribution is
\begin{equation}
\mathbf{P}_m^{\mathrm{elec}}
=
\epsilon_m J_m
\left[
\mathbf{e}_m\otimes
\left(\mathbf{C}_m^{-1}\mathbf{E}_m\right)
-
\frac{1}{2}
(\mathbf{e}_m\cdot\mathbf{e}_m)
\mathbf{F}_m^{-T}
\right].
\label{eq:matrix-electric-piola}
\end{equation}
Combining the mechanical and electrical contributions gives
\begin{equation}
\mathbf{P}_m
=
\mathbf{P}_m^{\mathrm{mech}}
+
\mathbf{P}_m^{\mathrm{elec}},
\qquad
\boldsymbol{\sigma}_m
=
J_m^{-1}\mathbf{P}_m\mathbf{F}_m^T,
\label{eq:matrix-total-stress}
\end{equation}
with the spatial electric contribution
\begin{equation}
\boldsymbol{\sigma}_m^{\mathrm{elec}}
=
\epsilon_m
\left[
\mathbf{e}_m\otimes\mathbf{e}_m
-
\frac{1}{2}
(\mathbf{e}_m\cdot\mathbf{e}_m)\mathbf{I}
\right].
\label{eq:matrix-maxwell-stress}
\end{equation}
Thus, deformation affects the dielectric response through
$J_m\mathbf{C}_m^{-1}$, while the electric field contributes directly to
the matrix stress. This matrix electromechanical response is combined with
the reduced electroactive beam model introduced next.

\subsection{Electroactive Beam Kinematics and Section Energy}
\label{sec:beam-electromechanical-model}

%
The embedded fibers are represented by geometrically exact beams following
the centerline--rotation description of \cite{Simo1985,SimoVuQuoc1988}.
Let $\mathbf r_0(s)$ denote the reference centerline and
$\mathbf Q_0(s)=[\mathbf e_t(s)\;\mathbf e_n(s)\;\mathbf e_b(s)]$ the
corresponding reference section basis, where $s$ is the reference arclength
coordinate and $t$, $n$, and $b$ denote the local tangent, normal, and
binormal directions. The mechanical state is described by the generalized
strain and curvature vectors
\begin{equation}
\boldsymbol{\gamma}
=
(\gamma_t,\gamma_n,\gamma_b),
\qquad
\boldsymbol{\kappa}
=
(\kappa_t,\kappa_n,\kappa_b),
\label{eq:beam-generalized-state}
\end{equation}
which contain the axial and shear strains and the twist and bending
curvatures, respectively. The rotation-vector field introduced in
Sec.~\ref{sec:theoretical-setting} is stored in the fixed global basis and
is denoted here by
$\boldsymbol{\alpha}_g(s)\equiv\boldsymbol{\theta}_b(s)$. Its
representation in the reference section basis and the associated rotation
tensors are
\begin{equation}
\boldsymbol{\alpha}_\ell
=
\mathbf Q_0^T\boldsymbol{\alpha}_g,
\qquad
\mathbf R_g
=
\exp([\boldsymbol{\alpha}_g]_\times),
\qquad
\mathbf R_\ell
=
\exp([\boldsymbol{\alpha}_\ell]_\times)
=
\mathbf Q_0^T\mathbf R_g\mathbf Q_0 .
\label{eq:rotation-frame-map}
\end{equation}
Here, $[\mathbf a]_\times$ denotes the skew-symmetric tensor satisfying
$[\mathbf a]_\times\mathbf b=\mathbf a\times\mathbf b$, and
$\operatorname{axial}(\cdot)$ denotes the inverse mapping from a
skew-symmetric tensor to its axial vector.
With
$\mathbf u^{\mathrm{loc}}=\mathbf Q_0^T\mathbf u_b$ and
$\mathbf e_t^{\mathrm{loc}}=(1,0,0)^T$, the current centerline and section
orientation are
\begin{equation}
\mathbf r
=
\mathbf r_0+\mathbf Q_0\mathbf u^{\mathrm{loc}},
\qquad
\mathbf Q
=
\mathbf R_g\mathbf Q_0
=
\mathbf Q_0\mathbf R_\ell,
\qquad
\boldsymbol{\kappa}_0
=
\operatorname{axial}(\mathbf Q_0^T\mathbf Q_0'),
\end{equation}
and the generalized strain follows as
\begin{equation}
\boldsymbol{\gamma}
=
\mathbf R_\ell^T
\left(
\mathbf e_t^{\mathrm{loc}}
+
(\mathbf u^{\mathrm{loc}})'
+
\boldsymbol{\kappa}_0\times\mathbf u^{\mathrm{loc}}
\right)
-
\mathbf e_t^{\mathrm{loc}}.
\label{eq:beam-generalized-strain}
\end{equation}
Here and below, a prime denotes differentiation with respect to reference
arclength. For curved or polygonal reference centerlines, the change of the
reference basis must also enter the rotation derivative. Accordingly,
\begin{equation}
\boldsymbol{\alpha}_\ell'
=
\mathbf Q_0^T\boldsymbol{\alpha}_g'
-
\boldsymbol{\kappa}_0\times\boldsymbol{\alpha}_\ell,
\end{equation}
and the nonlinear curvature is
\begin{equation}
\boldsymbol{\kappa}
=
\mathbf R_\ell^T\boldsymbol{\kappa}_0
-
\boldsymbol{\kappa}_0
+
\operatorname{axial}
\left(
\mathbf R_\ell^T\mathbf R_\ell'
\right).
\label{eq:beam-generalized-curvature}
\end{equation}
The rotation derivative is evaluated using the exact right Jacobian of the
exponential map, with its small-angle expansion used near zero. These
definitions recover zero generalized strain in the reference configuration
and are invariant under a superposed rigid-body motion.

The generalized beam measures determine the deformation of each material
point of the reduced cross-section. For local section coordinates
$(x_n,x_b)$, the section strain vector is
\begin{equation}
\mathbf{a}(x_n,x_b)
=
\begin{bmatrix}
\gamma_t+\kappa_n x_b-\kappa_b x_n\\
\gamma_n-\kappa_t x_b\\
\gamma_b+\kappa_t x_n
\end{bmatrix}.
\label{eq:section-a-vector}
\end{equation}
Defining the local axial stretch by $\lambda_s=1+a_t>0$, we adopt the
incompressible equal-transverse-stretch section reduction
\begin{equation}
\mathbf F_b
=
\begin{bmatrix}
\lambda_s & 0 & 0\\
a_n & \lambda_s^{-1/2} & 0\\
a_b & 0 & \lambda_s^{-1/2}
\end{bmatrix},
\qquad
J_b=1,
\qquad
\mathbf C_b=\mathbf F_b^T\mathbf F_b,
\qquad
I_{1b}=\mathrm{tr}\,\mathbf C_b,
\label{eq:beam-section-F}
\end{equation}
so that
$I_{1b}=\lambda_s^2+a_n^2+a_b^2+2/\lambda_s$. This reduction imposes
incompressibility kinematically rather than solving for independent
transverse stretches. Consequently, independent cross-section warping and
transverse expansion modes are not represented, although the imposed
lateral stretch varies over the section under bending. Using these section
kinematics, the electroactive beam energy is taken as the reduced
ideal-dielectric enthalpy
\begin{equation}
\psi_b
=
\underbrace{
\frac{1}{2}\mu_b(I_{1b}-3)
}_{\psi_b^{\mathrm{mech}}}
-
\underbrace{
\frac{1}{2}\epsilon_e J_b\,
\mathbf E_b\cdot\mathbf C_b^{-1}\mathbf E_b
}_{\psi_b^{\mathrm{elec}}},
\label{eq:dea-density}
\end{equation}
where $\mathbf E_b$ denotes the referential electric field acting on the
beam section and $\epsilon_e$ is the beam dielectric coefficient. Its
construction directly from the surrounding matrix field is the central
mixed-dimensional electrical coupling and is introduced in
Sec.~\ref{sec:matrix-projected-electric-coupling}. The mechanical
coefficient is parameterized as
\begin{equation}
\mu_b
=
\frac{Y_b}{2(1+\nu_b)},
\label{eq:lame-beam}
\end{equation}
where $Y_b$ and $\nu_b$ are used to specify the beam shear modulus;
$\nu_b$ does not introduce a volumetric penalty because the section
kinematics already enforce $J_b=1$. Following the cross-section-integration
viewpoint used for electroactive beam models
\cite{HuangLeyendecker2022}, the complete beam contribution is obtained by
integrating the section energy over the reference cross-section $A_b$ and
the embedded centerline network,
\begin{equation}
\Pi_b
=
\int_{\Gamma_0}
\int_{A_b}
\psi_b\,dA\,ds.
\label{eq:beam-energy-integral}
\end{equation}
For a circular reference cross-section of radius $r$, the section integral is
evaluated by introducing the area coordinate $u=(\rho/r)^2$ and combining
Gauss--Legendre integration in $u$ with periodic trapezoidal integration in
the polar angle. The implemented rule is
\begin{equation}
\begin{aligned}
n_u&=3,
&
n_\theta&=12,
\\
\left(\xi_i,\omega_i\right)_{i=0}^{n_u-1}
&=
\left\{
\left(-\sqrt{\frac35},\frac59\right),
\left(0,\frac89\right),
\left(\sqrt{\frac35},\frac59\right)
\right\},
&&
\\
u_i
&=
\frac{1+\xi_i}{2},
&
\rho_i
&=
 r\sqrt{u_i},
\\
\theta_j
&=
\left(j+\tfrac12\right)\frac{2\pi}{n_\theta},
&
w_{ij}
&=
\frac{r^2\omega_i}{4}\frac{2\pi}{n_\theta},
\\
i&=0,1,2,
&
j&=0,\ldots,11 .
\end{aligned}
\label{eq:section-quadrature-rule}
\end{equation}
Because $\rho\,d\rho=(r^2/2)\,du$, the transformed weights include the polar
Jacobian without a singular endpoint. The resulting thirty-six-point rule
integrates Cartesian polynomial moments over the disk through total degree
ten exactly to roundoff, including the section area, centroidal first moments,
and quadratic bending moments. The same points and weights are used in the
beam residual, tangent, electric coupling, and postprocessed section fields.

\subsection{Mixed-Dimensional Beam--Matrix Coupling}
\label{sec:mixed-dimensional-coupling}

The matrix and beam models introduced above are combined as an overlapping
mixed-dimensional reinforcement formulation. Before imposing mechanical
compatibility, the matrix and beam contributions define the phase potential
\begin{equation}
\Pi_0
=
\Pi_m^{\mathrm{mech}}
+
\Pi_m^{\mathrm{elec}}
+
\Pi_b .
\label{eq:phase-potential}
\end{equation}
The matrix energy is integrated over $\Omega_{m0}$, while the beam energy is
integrated over the embedded centerlines $\Gamma_0$ and their assigned
cross-sections. Consequently, the reduced fibers are superposed on the
matrix rather than replacing a corresponding three-dimensional matrix
volume. For a constant beam cross-sectional area $A_b$, the associated
nominal geometric fiber fraction is
\begin{equation}
f_b^{\mathrm{nom}}
=
\frac{A_bL_\Gamma}{|\Omega_0|},
\qquad
L_\Gamma
=
\int_{\Gamma_0} ds .
\label{eq:nominal-fiber-volume-fraction}
\end{equation}
Thus, $f_b^{\mathrm{nom}}$ measures the volume associated with the embedded
beam geometry but does not define a partition of $\Omega_0$ into disjoint
matrix and fiber phases. In particular, the formulation represents
reinforcement added to the matrix rather than a phase-replacement
composite. As a consequence, assigning identical constitutive coefficients
to the matrix and beam does not recover the response of a homogeneous
material. For example, in an undeformed state with a uniform electric field
and matched dielectric coefficient $\epsilon$, the additive model gives the
trial response
\begin{equation}
\mathbf D_{\mathrm{trial}}
=
\epsilon\left(1+f_b^{\mathrm{nom}}\right)\mathbf E,
\label{eq:additive-identical-coefficient-limit}
\end{equation}
rather than $\epsilon\mathbf E$. This distinction follows directly from
the superposed beam energy and is retained throughout the present
formulation.

Within this additive representation, the matrix and beam are coupled
electrically through the dependence of the beam dielectric energy on the
matrix electric field (Sec.~\ref{sec:matrix-projected-electric-coupling})
and mechanically through the projected mortar constraint
(Sec.~\ref{sec:coupling-energy}).

\subsubsection{Matrix-to-Beam Electric Coupling}
\label{sec:matrix-projected-electric-coupling}


The electrical interaction between the three-dimensional matrix and the
reduced beam network is introduced by evaluating the matrix electric field
along the embedded beam centerlines. Let $\mathbf X_{\mathrm{ep}}(s)$
denote the reference position at which a beam quadrature point is embedded
in the matrix. The matrix field evaluated at this point and its
representation in the reference beam basis are
\begin{equation}
\mathbf E_{\mathrm{proj}}(s)
=
\mathbf E_m\!\left(\mathbf X_{\mathrm{ep}}(s)\right),
\qquad
\mathbf E_b(s)
=
\mathbf Q_0^T(s)\mathbf E_{\mathrm{proj}}(s).
\label{eq:projected-electric-field}
\end{equation}
Because $\mathbf E_m$ is a referential electric field, the fixed reference
basis $\mathbf Q_0$ introduced in
Sec.~\ref{sec:beam-electromechanical-model} is used for this transformation,
rather than the current director basis. The resulting $\mathbf E_b$ is the
field entering the beam dielectric enthalpy,
\begin{equation}
\psi_b^{\mathrm{elec}}
=
-\frac{1}{2}\epsilon_e J_b\,
\mathbf E_b\cdot\mathbf C_b^{-1}\mathbf E_b,
\qquad
\Pi_b^{\mathrm{elec}}
=
\int_{\Gamma_0}\int_{A_b}
\psi_b^{\mathrm{elec}}\,dA\,ds .
\label{eq:projected-beam-electric-energy}
\end{equation}
Because $\psi_b^{\mathrm{elec}}$ depends on both the matrix potential through
$\mathbf E_b$ and the beam kinematics through $J_b\mathbf C_b^{-1}$, its
variation contributes to both the matrix electrical equilibrium and the beam
mechanical equilibrium. After finite-element discretization, this coupling
generates the corresponding mixed electromechanical blocks of the monolithic
residual and tangent described in
Sec.~\ref{sec:stationarity-nonlinear-solve}.

In the discrete formulation, $\mathbf E_{\mathrm{proj}}$ is evaluated from
the matrix finite-element field on beam subsegments divided at
matrix-element intersections. The value obtained at the centerline is used
for every cross-section quadrature point associated with that beam location,
so the present coupling does not perform finite-radius spatial averaging of
the matrix field. When an embedded point lies on an interface shared by more
than one matrix element, one containing element is selected and its field
gradient is used; gradients from neighboring elements are not averaged.
Accordingly, the centerline field represents a finite-resolution sampling of
the matrix electric field. At the continuum level, a general $H^1$
potential does not provide a well-defined gradient trace on a
one-dimensional set in three dimensions, and the sensitivity of this
sampling to matrix-mesh resolution and relative beam--matrix-mesh placement
is therefore assessed separately in the numerical verification studies. The electric displacement conjugate to the
local beam field follows from the same enthalpy,
\begin{equation}
\mathbf D_b^{\mathrm{loc}}
=
-\frac{\partial\psi_b^{\mathrm{elec}}}{\partial\mathbf E_b}
=
\epsilon_e J_b\mathbf C_b^{-1}\mathbf E_b .
\label{eq:local-projected-D}
\end{equation}
Integrating this quantity over the reference cross-section and rotating it
back to the global reference basis gives
\begin{equation}
\mathbf D_b^{\mathrm{sec,loc}}
=
\int_{A_b}
\mathbf D_b^{\mathrm{loc}}\,dA,
\qquad
\mathbf D_b^{\mathrm{sec}}
=
\mathbf Q_0\mathbf D_b^{\mathrm{sec,loc}} .
\label{eq:projected-D-output}
\end{equation}
The section-integrated quantity $\mathbf D_b^{\mathrm{sec}}$ later
contributes to the homogenized electric displacement of the RVE. This
construction differs from electroactive beam formulations that introduce
independent beam electric-potential degrees of freedom and a cross-section
potential expansion \cite{HuangLeyendecker2022}; here the reduced beam
obtains its electric field directly from the surrounding
three-dimensional potential field.

\subsubsection{Embedded Beam--Matrix Mechanical Coupling}
\label{sec:coupling-energy}

%

Mechanical compatibility between the independently discretized beam and
matrix fields is enforced through a projected mortar constraint. The
construction follows mixed-dimensional and embedded beam--solid coupling
approaches \cite{mccune2000mixed,ShimMonaghanArmstrong2002,
CunhaBarrosSenaCruz2012,steinbrecher2020mortar}. For a scalar mortar row
$r$, the projected displacement gap is written in terms of the beam-side
and matrix-side projection operators as
\begin{equation}
g_r
=
\mathbf D_r\mathbf u_b
-
\mathbf M_r\mathbf u_m .
\label{eq:translational-gap-general}
\end{equation}
Here, $\mathbf D_r$ and $\mathbf M_r$ project the beam and matrix
displacement fields onto the same one-dimensional mortar space. Thus,
$g_r=0$ expresses compatibility in a weak projected sense rather than
requiring coincident beam and matrix nodes. When the formulation is
specialized to periodic homogenization, the matrix displacement is
decomposed into affine and periodic parts. The same constraint then
contains an additional known affine contribution $a_r$, as introduced in
Sec.~\ref{sec:boundary-load-control}.
The projected compatibility condition is enforced in two numerical forms.
Exact Lagrange-multiplier constraints are used for smaller structured
problems and verification studies, where the additional multiplier system
remains manageable and compatibility can be imposed without a penalty
parameter. For the large irregular networks, the multiplier unknowns and
resulting saddle-point system substantially increase the algebraic problem
size and require treatment of dependent constraint rows. We therefore use a
quadratic penalty formulation for these cases, which eliminates the
multiplier degrees of freedom at the cost of a small finite compatibility
gap. Both forms impose the same projected mechanical coupling and differ
only in how the compatibility condition is enforced numerically.

For penalty coupling, the contribution associated with row $r$ is
\begin{equation}
\Pi_{c,r}^{\mathrm{pen}}
=
\frac{1}{2}\frac{p}{s_r}g_r^2,
\qquad
\lambda_r^{\mathrm{pen}}
=
\frac{p}{s_r}g_r,
\label{eq:penalty-coupling-energy}
\end{equation}
where $p$ is the coupling penalty, $s_r$ is a reference-length scaling
associated with the mortar row, and
$\lambda_r^{\mathrm{pen}}$ is the energetic quantity conjugate to the
projected gap. The row scaling accounts for variations in the support of
the mortar basis and is defined from the corresponding raw support measure
$s_r^{\mathrm{raw}}$ as
\begin{equation}
s_r
=
\max\!\left(
s_r^{\mathrm{raw}},
\alpha_s\overline{s}^{\mathrm{raw}}
\right),
\qquad
\overline{s}^{\mathrm{raw}}
=
\frac{1}{N_{\mathrm{act}}}
\sum_{r\in\mathcal A}s_r^{\mathrm{raw}},
\qquad
\alpha_s=1,
\label{eq:mortar-row-scaling-floor}
\end{equation}
where $\mathcal A$ denotes the set of active mortar rows and
$N_{\mathrm{act}}=|\mathcal A|$. With $s_r$ having dimensions of length
and $g_r$ dimensions of length squared, $p$ has units of stress and
$(p/s_r)g_r^2$ has units of energy. Summing over all active rows gives
\begin{equation}
\Pi_c^{\mathrm{pen}}
=
\sum_{r\in\mathcal A}
\Pi_{c,r}^{\mathrm{pen}} .
\label{eq:total-penalty-coupling-energy}
\end{equation}
Alternatively, exact translational compatibility is imposed by adding the
constraint contribution
\begin{equation}
\Pi_c^{\mathrm{exact}}
=
\sum_{r\in\mathcal A}\lambda_r g_r ,
\label{eq:exact-mortar-coupling}
\end{equation}
where $\lambda_r$ is an independent Lagrange multiplier and stationarity
with respect to $\lambda_r$ enforces $g_r=0$. The penalty and exact forms
therefore represent two numerical treatments of the same projected
compatibility condition rather than different mechanical models. The
finite-element construction of $\mathbf D_r$ and $\mathbf M_r$, the
resulting nonlinear systems, and the compatibility measures used to assess
the two treatments are given in
Sec.~\ref{sec:stationarity-nonlinear-solve}.

\subsection{Finite-Element Discretization and Nonlinear Solution}
\label{sec:stationarity-nonlinear-solve}

%
%
%
%
%
%

The matrix displacement and electric potential are discretized with
continuous quadratic tetrahedral shape functions $N_A(\mathbf X)$ on a
straight-sided linear geometry mapping. Linear interpolation is retained
as a comparison case in the numerical verification studies. Beam
translations and rotation vectors are interpolated linearly along each
centerline element using shape functions $N_a^b(s)$, while the mortar
space uses linear multiplier functions $N_r^\lambda(s)$ without bubble
enrichment. Accordingly, the discrete matrix and beam fields are
\begin{equation}
\mathbf u_m^h(\mathbf X)
=
\sum_A N_A(\mathbf X)\mathbf U_A,
\qquad
\Phi^h(\mathbf X)
=
\sum_A N_A(\mathbf X)\Phi_A,
\label{eq:matrix-discrete-fields}
\end{equation}
and
\begin{equation}
\mathbf u_b^h(s)
=
\sum_a N_a^b(s)\mathbf d_a,
\qquad
\boldsymbol{\theta}_b^h(s)
=
\sum_a N_a^b(s)\boldsymbol{\vartheta}_a,
\label{eq:beam-discrete-fields}
\end{equation}
where $\mathbf U_A$ and $\Phi_A$ are matrix displacement and electric-potential
coefficients, and $\mathbf d_a$ and $\boldsymbol{\vartheta}_a$ are beam
translation and global-basis rotation coefficients. The latter interpolate
the global rotation-vector field $\boldsymbol{\alpha}_g$ introduced in
Sec.~\ref{sec:beam-electromechanical-model}. Collecting the primal
coefficients gives
\begin{equation}
\mathbf q
=
\{\mathbf U_A,\Phi_A,\mathbf d_a,\boldsymbol{\vartheta}_a\}.
\label{eq:global-unknown-vector}
\end{equation}
For the periodic calculations introduced later, the same matrix spaces are
applied to the periodic fluctuation fields after separation of the prescribed
affine parts. Matrix integration uses quadrature order four and embedded beam–matrix integration uses eight Gauss points on each centerline subsegment contained
within a matrix element. The mortar projection operators introduced in
Sec.~\ref{sec:coupling-energy} are assembled on the same embedded
centerline geometry as
\begin{equation}
\begin{aligned}
    D_{ra}
&=
\sum_q
N_r^\lambda(s_q)N_a^b(s_q)J_qw_q,
&
M_{rA}
&=
\sum_q
N_r^\lambda(s_q)N_A(\mathbf X_{\mathrm{ep},q})J_qw_q,
\\
s_r^{\mathrm{raw}}
&=
\sum_q
N_r^\lambda(s_q)J_qw_q,
\end{aligned}
\end{equation}\label{eq:mortar-row-scaling-raw}

where $q$ indexes the embedded quadrature points and $J_qw_q$ is the
reference-centerline quadrature measure. The raw support
$s_r^{\mathrm{raw}}$ enters the scaled quantity $s_r$ defined in
Eq.~\eqref{eq:mortar-row-scaling-floor}. With these approximations, the
matrix, beam, projected-electric, and mechanical-coupling terms are
assembled into a single algebraic system. In particular, because the beam
electric energy depends simultaneously on the matrix electric-potential
coefficients through $\mathbf E_b$ and on the beam coefficients through
the section kinematics, its discrete derivatives
\begin{equation}
R_i^{\mathrm{elec},b}
=
\frac{\partial\Pi_b^{\mathrm{elec}}}{\partial q_i},
\qquad
K_{ij}^{\mathrm{elec},b}
=
\frac{\partial^2\Pi_b^{\mathrm{elec}}}
{\partial q_i\partial q_j}
\label{eq:projected-residual-tangent}
\end{equation}
generate both diagonal and mixed electromechanical blocks of the monolithic
residual and tangent.

For penalty coupling, the discrete potential is
\begin{equation}
\Pi_h^{\mathrm{pen}}(\mathbf q)
=
\Pi_0(\mathbf q)
+
\Pi_c^{\mathrm{pen}}(\mathbf q),
\end{equation}
where $\Pi_0$ is the matrix-plus-beam phase potential defined in
Sec.~\ref{sec:mixed-dimensional-coupling}. The nonlinear equilibrium
equations and consistent tangent are therefore
\begin{equation}
\mathbf R(\mathbf q)
=
\frac{\partial\Pi_h^{\mathrm{pen}}}{\partial\mathbf q}
=
\mathbf 0,
\qquad
K_{ij}
=
\frac{\partial R_i}{\partial q_j}
=
\frac{\partial^2\Pi_h^{\mathrm{pen}}}
{\partial q_i\partial q_j}.
\label{eq:stationarity}
\end{equation}
For the linear projected gap used here, the penalty contribution takes the
form
\begin{equation}
R_i^c
=
\sum_{r\in\mathcal A}
\frac{p}{s_r}g_r
\frac{\partial g_r}{\partial q_i},
\qquad
K_{ij}^c
=
\sum_{r\in\mathcal A}
\frac{p}{s_r}
\frac{\partial g_r}{\partial q_i}
\frac{\partial g_r}{\partial q_j}.
\label{eq:penalty-residual-gap}
\end{equation}
No independent multiplier equation is present in this case, and the
remaining finite gap is monitored separately from equilibrium convergence.
For exact translational coupling, the multiplier coefficients are added as
independent unknowns. Writing $\mathbf g$ for the vector of active projected
constraints, $\boldsymbol{\lambda}$ for the corresponding multipliers, and
$\mathbf B=\partial\mathbf g/\partial\mathbf q$, the discrete Lagrangian and
Newton system are
\begin{equation}
\mathcal L_h
=
\Pi_0+\boldsymbol{\lambda}^T\mathbf g,
\qquad
\mathbf z
=
(\mathbf q,\boldsymbol{\lambda}),
\end{equation}
and
\begin{equation}
\begin{bmatrix}
\mathcal L_{,qq} & \mathbf B^T\\
\mathbf B & \mathbf 0
\end{bmatrix}
\begin{bmatrix}
\Delta\mathbf q\\
\Delta\boldsymbol{\lambda}
\end{bmatrix}
=
-
\begin{bmatrix}
\Pi_{0,q}+\mathbf B^T\boldsymbol{\lambda}\\
\mathbf g
\end{bmatrix}.
\label{eq:exact-mortar-system}
\end{equation}
Thus the exact formulation enforces $\mathbf g=\mathbf 0$ through a
saddle-point problem, whereas the penalty formulation retains only the
primal unknowns and permits a small finite gap. As discussed in
Sec.~\ref{sec:coupling-energy}, the exact form is used for smaller
structured problems and verification studies, while the penalty form is
used for the large irregular networks where introducing a multiplier for
every retained constraint would substantially increase the algebraic
system. Exact-mode solves are initialized with temporary penalty
continuation stages before switching to the unpenalized KKT system; the
penalty parameter used during this continuation does not enter the final
exact constraint equations.

In both coupling modes, the nonlinear problem is advanced with load
prediction, residual-based line search, and adaptive load increments.
Convergence is assessed from the residuals associated with the beam and
matrix unknowns, and the projected beam--matrix gap is additionally
monitored for penalty coupling. Empty mortar rows are removed before
constraint condensation, and linearly dependent constraint rows are removed
for the exact formulation. Trial states are accepted only when $J_m>0$ at all matrix quadrature points
and $\lambda_s>0$ at all active beam cross-section points.
Network junctions are treated directly at the discrete level. Structured
center-cross networks share both translational and rotation-vector degrees
of freedom at a common junction, producing a rigid crosslink. Voronoi
branches share translations but retain independent rotations and therefore
act as rotational hinges. No rotational beam--matrix mortar constraint is
imposed in the calculations reported here.

\subsection{Periodic Computational Homogenization}
\label{sec:homogenization-background}


We now specialize the mixed-dimensional formulation to periodic
representative-volume-element calculations. Following first-order
computational homogenization at finite strain
\cite{kouznetsova2001micro,keip2014twoscale}, the macroscopic loading is
described by the deformation gradient $\mathbf F_{\mathrm{macro}}$ and
referential electric field $\mathbf E_{\mathrm{macro}}$. These quantities
define the affine parts of the matrix displacement and electric potential,
while the remaining microscopic fields are represented by the periodic
fluctuations $\widetilde{\mathbf u}_m$ and $\widetilde{\phi}$ together with
the beam displacement and rotation fields. The precise affine-plus-periodic
decompositions, the corresponding beam periodicity conditions, and the
loading paths used in the numerical examples are given in
Sec.~\ref{sec:boundary-load-control}.

For each prescribed pair
$(\mathbf F_{\mathrm{macro}},\mathbf E_{\mathrm{macro}})$, the microscopic
electromechanical problem is solved to equilibrium subject to these periodic
constraints. The resulting stationary cell response defines the homogenized
first Piola--Kirchhoff stress
$\overline{\mathbf P}_{\mathrm{RVE}}$ and referential electric displacement
$\overline{\mathbf D}_{\mathrm{RVE}}$. Importantly, these effective quantities
are obtained from variations of the complete coupled cell potential with
respect to the macroscopic loading variables, so the matrix, beam,
matrix-to-beam electrical contribution, and mechanical coupling work all
enter the micro-to-macro response. Their explicit definitions are given in
Sec.~\ref{sec:rve-effective-quantities}, while the associated energetic
consistency and macro-homogeneity condition are established in
Sec.~\ref{sec:macro-homogeneity}.

\subsubsection{Periodic Boundary Conditions and Load Control}
\label{sec:boundary-load-control}

%
%
%
%
%
Periodic mechanical loading is introduced by decomposing the total matrix
displacement into a prescribed affine part and a periodic fluctuation,
\begin{equation}
\mathbf u_m(\mathbf X)
=
(\mathbf F_{\mathrm{macro}}-\mathbf I)\mathbf X
+
\widetilde{\mathbf u}_m(\mathbf X),
\qquad
\widetilde{\mathbf u}_m(\mathbf X^+)
=
\widetilde{\mathbf u}_m(\mathbf X^-),
\label{eq:rve-displacement-decomposition}
\end{equation}
where $\mathbf X^+$ and $\mathbf X^-$ denote paired points on opposite
faces of the reference cell. Consequently, the total displacement jump is
\begin{equation}
\mathbf u_m(\mathbf X^+)
-
\mathbf u_m(\mathbf X^-)
=
(\mathbf F_{\mathrm{macro}}-\mathbf I)
(\mathbf X^+-\mathbf X^-),
\label{eq:periodic-mechanical-bc}
\end{equation}
and the local matrix deformation gradient becomes
\begin{equation}
\mathbf F_m(\mathbf X)
=
\mathbf F_{\mathrm{macro}}
+
\nabla_{\!X}\widetilde{\mathbf u}_m(\mathbf X).
\label{eq:rve-F-from-fluctuation}
\end{equation}
The embedded beams satisfy the corresponding periodic displacement map. For
paired beam endpoints with reference separation
$\Delta\mathbf X=\mathbf X^+-\mathbf X^-$,
\begin{equation}
\mathbf u_b^+
-
\mathbf u_b^-
=
(\mathbf F_{\mathrm{macro}}-\mathbf I)\Delta\mathbf X,
\qquad
\boldsymbol{\vartheta}_b^+
=
\boldsymbol{\vartheta}_b^-,
\label{eq:beam-periodic-map}
\end{equation}
where the rotation vectors are identified componentwise in the fixed global
basis. With the matrix field written in terms of
$\widetilde{\mathbf u}_m$, the general mortar constraint from
Sec.~\ref{sec:coupling-energy} takes the RVE-specific form
\begin{equation}
g_r
=
\mathbf D_r\mathbf u_b
-
\mathbf M_r\widetilde{\mathbf u}_m
+
a_r ,
\label{eq:translational-gap}
\end{equation}
where $a_r$ is the known contribution generated by the prescribed affine
matrix displacement, including the periodic-image correction for beam segments crossing an RVE boundary. The constraint therefore continues to compare total
beam and matrix displacements even though only the matrix fluctuation field
appears explicitly in Eq.~\eqref{eq:translational-gap}. A displacement
anchor removes the remaining rigid-translation gauge. Beam rotations are
otherwise left free rather than pinned, since an artificial rotational
constraint would introduce reaction moments. For straight circular fibers with an unconstrained axial spin mode, a single constant-spin gauge removes this null mode.

The electrical boundary condition follows the same affine-plus-periodic
construction. The total matrix potential is written as
\begin{equation}
\Phi(\mathbf X)
=
-\mathbf E_{\mathrm{macro}}\cdot\mathbf X
+
\widetilde{\phi}(\mathbf X),
\qquad
\widetilde{\phi}(\mathbf X^+)
=
\widetilde{\phi}(\mathbf X^-),
\label{eq:electric-potential-periodic-split}
\end{equation}
so that
\begin{equation}
\Phi(\mathbf X^+)-\Phi(\mathbf X^-)
=
-\mathbf E_{\mathrm{macro}}\cdot
(\mathbf X^+-\mathbf X^-)
\label{eq:periodic-electric-potential-jump}
\end{equation}
and the local referential electric field is
\begin{equation}
\mathbf E_m(\mathbf X)
=
-\nabla_{\!X}\Phi
=
\mathbf E_{\mathrm{macro}}
-
\nabla_{\!X}\widetilde{\phi}(\mathbf X).
\label{eq:electric-field-periodic-bc}
\end{equation}
Because the beams carry no independent electric-potential field, no
additional electrical periodicity condition is required on the beam
network. A single scalar potential anchor removes the arbitrary additive
constant in $\widetilde{\phi}$ without altering the prescribed macroscopic
electric field. For an unperforated periodic cell,
$\Omega_{m0}=\Omega_0$, periodicity of the fluctuation fields gives the
average-field identities
\begin{equation}
\frac{1}{|\Omega_0|}
\int_{\Omega_{m0}}
\mathbf F_m\,dV
=
\mathbf F_{\mathrm{macro}},
\qquad
\frac{1}{|\Omega_0|}
\int_{\Omega_{m0}}
\mathbf E_m\,dV
=
\mathbf E_{\mathrm{macro}}.
\label{eq:periodic-average-fields}
\end{equation}
These identities provide useful kinematic and electrical consistency checks.
For a perforated cell, however, the internal void boundary contributes to
the fluctuation gradients. Defining
$f_m=|\Omega_{m0}|/|\Omega_0|$ and letting $\mathbf N_m$ denote the
reference normal directed outward from the matrix into the void gives
\begin{align}
\frac{1}{|\Omega_0|}
\int_{\Omega_{m0}}\mathbf F_m\,dV
&=
f_m\mathbf F_{\mathrm{macro}}
+
\frac{1}{|\Omega_0|}
\int_{\partial\Omega_{v0}}
\widetilde{\mathbf u}_m\otimes\mathbf N_m\,dA,
\\
\frac{1}{|\Omega_0|}
\int_{\Omega_{m0}}\mathbf E_m\,dV
&=
f_m\mathbf E_{\mathrm{macro}}
-
\frac{1}{|\Omega_0|}
\int_{\partial\Omega_{v0}}
\widetilde{\phi}\mathbf N_m\,dA.
\label{eq:perforated-average-fields}
\end{align}
The full-cell identities in Eq.~\eqref{eq:periodic-average-fields} therefore
do not apply directly to perforated matrix domains.

The loading paths are advanced quasi-statically using a scalar load
parameter $\ell\in[0,1]$. Unless otherwise specified, prescribed mechanical
components are interpolated linearly from the undeformed state to a target
macroscopic deformation $\mathbf F_\star$,
\begin{equation}
\mathbf F_{\mathrm{macro}}(\ell)
=
\mathbf I
+
\ell(\mathbf F_\star-\mathbf I),
\qquad
J_{\mathrm{macro}}(\ell)
=
\det\mathbf F_{\mathrm{macro}}(\ell).
\label{eq:actual-linear-load}
\end{equation}
Because this interpolation is componentwise, an isochoric endpoint does not
in general imply an isochoric loading path. Cases that enforce constant
volume throughout the loading history are therefore prescribed with an
explicitly isochoric path instead. Electrical loading is either advanced
with the same load parameter or held fixed throughout the mechanical
loading,
\begin{equation}
\mathbf E_{\mathrm{macro}}(\ell)
=
\begin{cases}
\ell\,\mathbf E_{\mathrm{target}},
&
\text{ramped electric loading},
\\[2mm]
\mathbf E_{\mathrm{target}},
&
\text{constant-bias loading},
\end{cases}
\label{eq:general-electric-load}
\end{equation}
where $\mathbf E_{\mathrm{target}}$ is the prescribed target referential
electric field. The constant-bias path is used for the sensing calculations,
so the electrical field is present from the initial state while the
mechanical deformation is subsequently varied.

\subsubsection{Effective RVE Stress and Electric Displacement}
\label{sec:rve-effective-quantities}

%
%
%
%

The homogenized electromechanical response is defined from the complete
discrete cell functional evaluated at microscopic equilibrium. Let
$\mathcal H_h$ denote the penalty potential $\Pi_h^{\mathrm{pen}}$ in
penalty mode and the discrete Lagrangian $\mathcal L_h$ in exact-mortar
mode. The effective first Piola--Kirchhoff stress and referential electric
displacement are the energetic quantities conjugate to the prescribed
macroscopic deformation gradient and electric field,
\begin{equation}
\overline{\mathbf P}_{\mathrm{RVE}}
=
\frac{1}{|\Omega_0|}
\frac{d\mathcal H_h}{d\mathbf F_{\mathrm{macro}}},
\qquad
\overline{\mathbf D}_{\mathrm{RVE}}
=
-\frac{1}{|\Omega_0|}
\frac{d\mathcal H_h}{d\mathbf E_{\mathrm{macro}}}.
\label{eq:rve-effective-conjugates}
\end{equation}
These derivatives follow the affine and periodic maps introduced in
in Sec.~\ref{sec:boundary-load-control} while the independent microscopic
fluctuation variables are held fixed. At equilibrium, their implicit
variations do not contribute because the microscopic stationarity
conditions are satisfied. Consequently, the derivatives in
Eq.~\eqref{eq:rve-effective-conjugates} contain the matrix and beam energies,
the matrix-to-beam electric enthalpy, and the macroscopic work
associated with the mechanical coupling. In particular, the mortar
contribution to the effective stress can be written compactly as
\begin{equation}
\overline{\mathbf P}_c
=
\frac{1}{|\Omega_0|}
\sum_{r\in\mathcal A}
\lambda_r
\frac{\partial g_r}{\partial\mathbf F_{\mathrm{macro}}},
\label{eq:coupling-macro-stress}
\end{equation}
where the derivative includes all dependence of the projected gap on the
prescribed affine displacement and periodic beam maps. Here,
$\lambda_r=(p/s_r)g_r$ in penalty mode, whereas $\lambda_r$ is the solved
Lagrange multiplier in exact-mortar mode. Thus, even when the exact
constraint satisfies $g_r=0$, its multiplier can contribute nonzero
macroscopic work. The effective Cauchy stress follows from the usual
finite-strain transformation
\begin{equation}
\overline{\boldsymbol{\sigma}}_{\mathrm{RVE}}
=
\frac{1}{J_{\mathrm{macro}}}
\overline{\mathbf P}_{\mathrm{RVE}}
\mathbf F_{\mathrm{macro}}^T,
\qquad
J_{\mathrm{macro}}
=
\det\mathbf F_{\mathrm{macro}} .
\label{eq:rve-cauchy-sum}
\end{equation}
No componentwise symmetrization is applied to this quantity.

For interpretation, matrix- and beam-only stress measures are also retained,
but they are diagnostics rather than definitions of the effective stress.
The matrix contribution is obtained by direct averaging of the local Cauchy
stress over the current matrix domain,
\begin{equation}
\overline{\boldsymbol{\sigma}}_m
=
\frac{1}{|\Omega|}
\int_{\Omega_m}
\boldsymbol{\sigma}_m\,dv,
\label{eq:matrix-rve-cauchy}
\end{equation}
where $|\Omega|$ is the current volume of the enclosing RVE. For the beam
network, let
$\{\mathbf d_1,\mathbf d_2,\mathbf d_3\}$ denote the current orthonormal
section directors, with $\mathbf d_1$ aligned with the beam axis, and let
$N$, $Q_n$, and $Q_b$ denote the axial and transverse section-force
resultants obtained from the beam energy. A symmetric stress-like section
resultant is reconstructed as
\begin{equation}
\mathbf K_b
=
N\,\mathbf d_1\otimes\mathbf d_1
+
Q_n
\left(
\mathbf d_1\otimes\mathbf d_2
+
\mathbf d_2\otimes\mathbf d_1
\right)
+
Q_b
\left(
\mathbf d_1\otimes\mathbf d_3
+
\mathbf d_3\otimes\mathbf d_1
\right),
\label{eq:beam-section-cauchy}
\end{equation}
which gives the corresponding beam diagnostic
\begin{equation}
\overline{\boldsymbol{\sigma}}_b
=
\frac{1}{|\Omega|}
\int_{\Gamma}
\mathbf K_b\,ds .
\label{eq:beam-rve-cauchy}
\end{equation}
Bending and torsional resultants describe the variation of stress across a
beam section but have zero section mean about the centroid and therefore do
not appear in Eq.~\eqref{eq:beam-section-cauchy}. Moreover, this
resultant-based reconstruction has not been identified with the integrated
Cauchy stress of an unrestricted three-dimensional fiber. For these reasons,
neither $\overline{\boldsymbol{\sigma}}_m+
\overline{\boldsymbol{\sigma}}_b$ nor either phase contribution separately
replaces the energetic effective stress in
Eq.~\eqref{eq:rve-effective-conjugates}.

The electric response admits a corresponding phase decomposition because the
mechanical mortar constraint carries no independent electric-potential
variable. The matrix contribution, normalized by the enclosing reference-cell
volume, is
\begin{equation}
\overline{\mathbf D}_m
=
\frac{1}{|\Omega_0|}
\int_{\Omega_{m0}}
\mathbf D_m\,dV
=
\frac{1}{|\Omega_0|}
\int_{\Omega_{m0}}
\epsilon_m J_m\mathbf C_m^{-1}\mathbf E_m\,dV .
\label{eq:matrix-electric-average}
\end{equation}
The embedded beams contribute through the section-integrated referential
electric displacement $\mathbf D_b^{\mathrm{sec}}$ defined in
Sec.~\ref{sec:matrix-projected-electric-coupling}. Hence,
\begin{equation}
\overline{\mathbf D}_{\mathrm{RVE}}
=
\overline{\mathbf D}_m
+
\frac{1}{|\Omega_0|}
\int_{\Gamma_0}
\mathbf D_b^{\mathrm{sec}}(s)\,ds .
\label{eq:rve-electric-displacement}
\end{equation}
Both contributions are therefore expressed per enclosing reference-cell
volume, consistent with the energetic definition in
Eq.~\eqref{eq:rve-effective-conjugates}. The relation of these effective
quantities to microscopic and macroscopic work is established next through
the macro-homogeneity condition.

\subsubsection{Macro-Homogeneity}
\label{sec:macro-homogeneity}

%
%

The effective quantities introduced above satisfy a discrete
macro-homogeneity condition because they are obtained from the stationary
cell functional. Let
\begin{equation}
\mathbf A
=
\left(
\mathbf F_{\mathrm{macro}},
\mathbf E_{\mathrm{macro}}
\right)
\end{equation}
collect the prescribed macroscopic variables, and let $\mathbf z$ denote
the microscopic algebraic state. In penalty mode,
$\mathbf z=\mathbf q$, whereas in exact-mortar mode
$\mathbf z=(\mathbf q,\boldsymbol{\lambda})$. Using the unified cell
functional $\mathcal H_h$ introduced in
Sec.~\ref{sec:rve-effective-quantities}, the stationary microscopic solution
for a prescribed macroscopic state satisfies
\begin{equation}
\frac{\partial\mathcal H_h}{\partial\mathbf z}
\left(
\mathbf z^\star(\mathbf A),\mathbf A
\right)
=
\mathbf 0 .
\label{eq:microscopic-stationarity-homogenization}
\end{equation}
The corresponding homogenized cell enthalpy is
\begin{equation}
\overline{\Psi}_h(\mathbf A)
=
\frac{1}{|\Omega_0|}
\mathcal H_h
\left(
\mathbf z^\star(\mathbf A),\mathbf A
\right).
\label{eq:homogenized-cell-enthalpy}
\end{equation}
Its variation along a differentiable stationary branch follows from the
chain rule,
\begin{align}
\delta\overline{\Psi}_h
&=
\frac{1}{|\Omega_0|}
\left[
\frac{\partial\mathcal H_h}{\partial\mathbf A}
\cdot\delta\mathbf A
+
\frac{\partial\mathcal H_h}{\partial\mathbf z}
\cdot\delta\mathbf z^\star
\right]
\nonumber\\
&=
\overline{\mathbf P}_{\mathrm{RVE}}
:
\delta\mathbf F_{\mathrm{macro}}
-
\overline{\mathbf D}_{\mathrm{RVE}}
\cdot
\delta\mathbf E_{\mathrm{macro}},
\label{eq:coupled-discrete-work}
\end{align}
where the second term in the first line vanishes by microscopic
stationarity. In exact-mortar mode this statement includes both primal
equilibrium and satisfaction of the multiplier equations
$\mathbf g=\mathbf 0$. The minus sign in the electrical term follows from
the fixed-potential enthalpy convention used throughout the formulation.

Equation~\eqref{eq:coupled-discrete-work} is the discrete
electromechanical Hill--Mandel condition for the mixed-dimensional RVE: the
macroscopic virtual work equals the change of the complete microscopic cell
functional per reference-cell volume. More explicitly, the macroscopic
variation can be decomposed as
\begin{equation}
|\Omega_0|
\left(
\overline{\mathbf P}_{\mathrm{RVE}}
:
\delta\mathbf F_{\mathrm{macro}}
-
\overline{\mathbf D}_{\mathrm{RVE}}
\cdot
\delta\mathbf E_{\mathrm{macro}}
\right)
=
\delta_{\mathbf A}\Pi_m^{\mathrm{mech}}
+
\delta_{\mathbf A}\Pi_m^{\mathrm{elec}}
+
\delta_{\mathbf A}\Pi_b
+
\delta_{\mathbf A}\Pi_c ,
\label{eq:mixed-dimensional-hill-mandel}
\end{equation}
where $\delta_{\mathbf A}$ denotes variation with respect to the prescribed
macroscopic fields while the independent microscopic coefficients are held
fixed. In penalty mode,
$\Pi_c=\Pi_c^{\mathrm{pen}}$, whereas in exact-mortar mode the coupling
contribution is
$\Pi_c=\boldsymbol{\lambda}^T\mathbf g$. Although
$\mathbf g=\mathbf 0$ at an exact-mortar solution, its variation with
respect to the macroscopic deformation need not vanish, so the multiplier
term contributes to the macroscopic mechanical work. The same identity also
includes the electrical work of the embedded beams through the dependence of
$\Pi_b$ on the matrix-to-beam electric field. Thus the effective response
accounts consistently for matrix, beam, electrical, and mechanical-coupling
contributions rather than being reconstructed from phase averages alone.

\subsection{Derived Quantities and Electrical Response Measures}
\label{sec:sensor-quantities}

%
%
%
%
%
%

In addition to the effective quantities defined in
Sec.~\ref{sec:rve-effective-quantities}, several derived measures are used
to assess the local deformation and to present the electromechanical
response. The matrix deformation averages are evaluated with respect to the
enclosing reference-cell volume,
\begin{equation}
\overline{\mathbf F}_m
=
\frac{1}{|\Omega_0|}
\int_{\Omega_{m0}}
\mathbf F_m\,dV,
\qquad
\overline{J}_m
=
\frac{1}{|\Omega_0|}
\int_{\Omega_{m0}}
J_m\,dV .
\label{eq:derived-F-J}
\end{equation}
For an unperforated cell with compatible periodic kinematics,
$\overline{\mathbf F}_m=\mathbf F_{\mathrm{macro}}$, while the average
Jacobian provides a global volume-change diagnostic. Neither quantity,
however, establishes local incompressibility. The latter is assessed from
the spatial distribution of $J_m$, including its extrema and deviations
from unity. As discussed in Sec.~\ref{sec:boundary-load-control}, these
full-cell identities are modified for perforated domains. When alternative
stress measures are required, they are obtained from the energetic RVE
stress through
\begin{equation}
\overline{\mathbf P}_{\mathrm{RVE}}
=
J_{\mathrm{macro}}
\overline{\boldsymbol{\sigma}}_{\mathrm{RVE}}
\mathbf F_{\mathrm{macro}}^{-T},
\qquad
\overline{\mathbf S}_{\mathrm{RVE}}
=
\mathbf F_{\mathrm{macro}}^{-1}
\overline{\mathbf P}_{\mathrm{RVE}},
\label{eq:derived-stress-measures}
\end{equation}
where $\overline{\mathbf S}_{\mathrm{RVE}}$ is the effective second
Piola--Kirchhoff stress. The effective Cauchy stress
$\overline{\boldsymbol{\sigma}}_{\mathrm{RVE}}$ remains the primary
mechanical output used in the numerical comparisons.

The primary electrical output is the homogenized referential electric
displacement $\overline{\mathbf D}_{\mathrm{RVE}}$. Depending on the loading
configuration, the results are reported either through individual
components or through its magnitude,
\begin{equation}
\left|
\overline{\mathbf D}_{\mathrm{RVE}}
\right|
=
\sqrt{
\overline{D}_{1,\mathrm{RVE}}^2
+
\overline{D}_{2,\mathrm{RVE}}^2
+
\overline{D}_{3,\mathrm{RVE}}^2
}.
\label{eq:rve-D-norm}
\end{equation}
For the fixed-bias calculations introduced in
Sec.~\ref{sec:boundary-load-control}, the electric field is applied before
the mechanical loading and then held constant while the RVE deforms. Let
$\eta$ denote the scalar coordinate parameterizing the prescribed mechanical
loading path and let $\eta_0$ denote the initially biased state. The electrical
response is characterized by
\begin{equation}
\Delta\overline{\mathbf D}_{\mathrm{RVE}}(\eta)
=
\overline{\mathbf D}_{\mathrm{RVE}}(\eta)
-
\overline{\mathbf D}_{\mathrm{RVE}}(\eta_0),
\qquad
\frac{
\Delta
\left|
\overline{\mathbf D}_{\mathrm{RVE}}
\right|
}{
\left|
\overline{\mathbf D}_{\mathrm{RVE}}
\right|_0
}
=
\frac{
\left|
\overline{\mathbf D}_{\mathrm{RVE}}(\eta)
\right|
-
\left|
\overline{\mathbf D}_{\mathrm{RVE}}(\eta_0)
\right|
}{
\left|
\overline{\mathbf D}_{\mathrm{RVE}}(\eta_0)
\right|
}.
\label{eq:sensor-rve-D-change}
\end{equation}
These quantities describe the change in electrical response produced by
mechanical deformation under a prescribed referential bias and are the
electrical response measures used in the numerical examples. If a specific electrode
configuration is assumed, the electrode-normal component of
$\overline{\mathbf D}_{\mathrm{RVE}}$ can additionally be related to free
charge, effective capacitance, and an estimated open-circuit voltage. That
conversion requires assumptions on electrode orientation, fringing,
leakage, and mechanical feedback during voltage readjustment and is
therefore treated separately as a postprocessing relation rather than as
part of the RVE solution.

\section{Numerical Verification and Model Assessment}
\label{sec:numerical-verification}

%

Before considering the response of different network architectures, we
examine the sensitivity of the mixed-dimensional calculations to the
numerical choices used throughout the study. We first summarize the common
material parameters and numerical discretization. A straight embedded fiber
is then used to assess the influence of matrix and beam resolution, followed
by a nonaffine curved-network problem to examine the matrix interpolation
order and J-bar volumetric treatment. The final studies focus on the
matrix-to-beam electric coupling by varying the matrix-mesh resolution and
the position of the beam network relative to that mesh. The corresponding
three-dimensional single-fiber comparison
and detailed numerical data are reported in the appendices.

\subsection{Material Parameters and Numerical Baseline}
\label{sec:baseline-coefficients}

%
%
%
%
%
The same constitutive parameters are used throughout the numerical studies
unless stated otherwise. The beam is assigned
$Y_b=1.0~\mathrm{MPa}$ and $\mu_b=0.334~\mathrm{MPa}$, representative of a
relatively stiff silicone dielectric elastomer, while the matrix uses
$Y_m=0.060~\mathrm{MPa}$ and $\mu_m=0.020~\mathrm{MPa}$. The resulting
beam-to-matrix stiffness ratio is approximately $16.7$. The nearly
incompressible matrix uses $C_{1m}=0.010~\mathrm{MPa}$ and
$K_m=2.0~\mathrm{MPa}$. The beam properties are based on reported values
for NuSil CF19-2186 \cite{banet2021evaluation}, while the matrix stiffness
is representative of an Ecoflex-like elastomer \cite{yu2015work}. These
parameters provide a common mechanically heterogeneous baseline rather
than a calibration to a particular fiber--matrix composite. The beam and
matrix are assigned the same dielectric coefficient,
$\epsilon_e=\epsilon_m=2.48\times10^{-5}$ in the MPa--MV/m unit system,
corresponding to a relative permittivity of approximately $2.8$. Values of
this magnitude have been reported for both NuSil CF19-2186 and Ecoflex-like
silicone elastomers \cite{banet2021evaluation,nikbakhtnasrabadi2021textile}.
The matched-permittivity choice avoids introducing an additional dielectric
contrast into the architecture study, so that differences in electrical
response arise from the network geometry, deformation, and
mixed-dimensional coupling. The applied referential electric fields considered below range
from $0$ to $50~\mathrm{MV/m}$. The complete parameter set is summarized in
Table~\ref{tab:latest-run-parameters}.

Unless stated explicitly, the matrix displacement and electric potential
are approximated with quadratic tetrahedral elements using the J-bar
volumetric treatment of Sec.~\ref{sec:matrix-electromechanical-model}. Beam translations
and rotations and the mortar interpolation are linear. Matrix and
beam--matrix quadrature orders are four and eight, respectively, and the
beam cross-section is integrated using the rule in
Eq.~\eqref{eq:section-quadrature-rule}. Exact translational mortar
constraints are used for the single-fiber and structured-network
calculations, where the multiplier systems remain moderate and provide a
direct enforcement of beam--matrix compatibility. The larger irregular
networks use penalty coupling to avoid the substantially larger
saddle-point systems associated with exact mortar. The sensitivity of the
results to the matrix and beam discretizations, the J-bar treatment, and
the sampling of the matrix electric field is examined in the following
subsections. Additional numerical settings and supporting data are given in
the appendices.

\subsection{Single-Fiber Discretization Sensitivity}
\label{sec:single-fiber-discretization}

%
%
%
%
%

\begin{figure}[htb]
    \centering
    \includegraphics[width=0.40\linewidth]{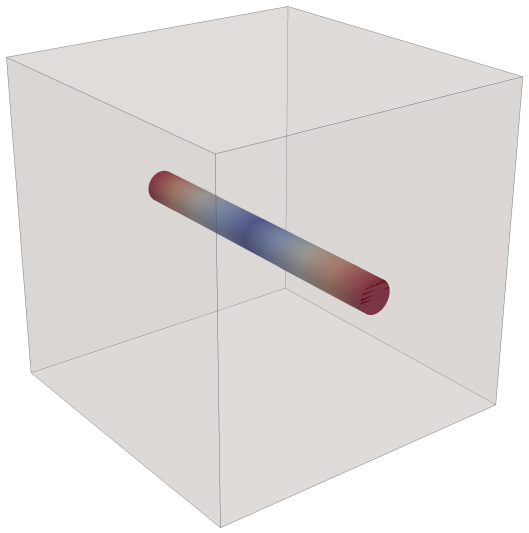}
    \caption{Single-fiber geometry used for the beam and matrix
    discretization study.}
    \label{fig:single_fiber_mesh}
\end{figure}

Because the beam and matrix are discretized independently, we first examine
how their relative resolution affects the mixed-dimensional response. A
single straight fiber provides a useful test because its mechanical response
is affine, without the spatial variations introduced by curvature or network
junctions. The fiber is embedded along the $x$ direction of a unit-cube
matrix and has length $L=1~\mathrm{m}$ and radius $r=0.05~\mathrm{m}$, as
shown in Fig.~\ref{fig:single_fiber_mesh}. The geometry and material
parameters are held fixed while the matrix mesh size $h_m$ and the number of
beam elements $N_b$ are varied. We define
\begin{equation}
    h_b=\frac{L}{N_b},
    \qquad
    \rho_h=\frac{h_b}{h_m},
    \label{eq:single-fiber-resolution-ratio}
\end{equation}
so that the study varies both the matrix resolution and the relative
resolution of the beam and matrix meshes. Exact translational mortar
coupling is used throughout.

For the mechanical calculation, the cell is stretched along the fiber
direction while the macroscopic volume is kept constant,
\begin{equation}
\begin{aligned}
\lambda(t)&=1+t(\lambda_\star-1), &
\eta(t)&=1+t(\lambda_\star^{-1/2}-1),\\
\mathbf F_{\mathrm{macro}}(t)
&=
\operatorname{diag}\left(
\lambda(t),\eta(t),[\lambda(t)\eta(t)]^{-1}
\right), &
\det\mathbf F_{\mathrm{macro}}(t)&=1 ,
\end{aligned}
\qquad t\in[0,1].
\label{eq:single-fiber-convergence-load}
\end{equation}
The embedded solution remains affine over the sampled beam and matrix
resolutions. The homogenized stress is unchanged to numerical precision,
with a maximum relative difference of $4.22\times10^{-10}$ from the
analytical affine response in
Eq.~\eqref{eq:single-fiber-affine-response}. The straight-fiber mechanical response is
therefore insensitive to the sampled discretizations. A comparison with a
body-fitted three-dimensional fiber is given in
Appendix~\ref{app:single-fiber-resolved-reference}. It is treated
separately because that comparison also includes the modeling difference
between the additive beam representation and a fiber that explicitly
replaces matrix material.

\begin{figure}[htb]
    \centering
    \includegraphics[width=0.72\linewidth]
    {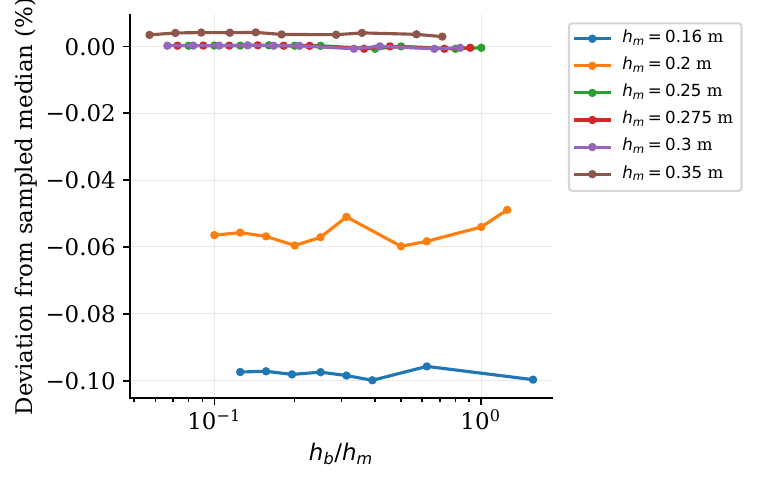}
    \caption{Variation of the final homogenized electric-displacement
    magnitude with matrix and beam resolution for the single-fiber
    electric-only calculation. Values are reported relative to the median
    response. }
    \label{fig:single-fiber-electric-D-sensitivity}
\end{figure}

To assess the discretization sensitivity of the electrical response, we use the same
single-fiber geometry in the undeformed configuration and apply a uniform macroscopic
electric field,
\begin{equation}
    \mathbf F_{\mathrm{macro}}=\mathbf I,
    \qquad
    \mathbf E_{\mathrm{macro}}
    =25\,\mathbf e_3~\mathrm{MV/m}.
    \label{eq:single-fiber-electric-convergence-load}
\end{equation}
Because the beam dielectric energy samples the matrix electric field along
the centerline, the electrical response can depend on the matrix
discretization even when the physical geometry is unchanged. Figure~\ref{fig:single-fiber-electric-D-sensitivity}
shows the variation in the homogenized electric-displacement
magnitude over the tested combinations of matrix and beam resolution. Across the
tested cases with $h_m\geq0.16~\mathrm{m}$, the absolute deviation from
the median response does not exceed $0.100\%$. Based on this sensitivity study, the structured examples use
$h_m=0.2~\mathrm{m}$ and eight beam elements per straight unit-length
fiber as the baseline discretization. Curved centerlines are discretized
more finely to resolve their geometry.

\subsection{Matrix Discretization and Volumetric Treatment}
\label{sec:higher-order-jbar}

%
%
%
%
%
%

The straight-fiber problem in Sec.~\ref{sec:single-fiber-discretization}
is affine and therefore provides only a limited test of the matrix
discretization. We next use the curved three-directional network, where
fiber curvature and the central junction generate a spatially varying
matrix deformation under a homogeneous macroscopic load. This makes the
response sensitive to both the interpolation of $\mathbf F_m$ and the
treatment of the volumetric term. It also makes the J-bar comparison
meaningful: for linear tetrahedra, $\mathbf F_m$ is constant within each
element and the element-averaged $\bar J_e$ coincides with the pointwise
determinant.

The beam discretization and mortar spaces are held fixed while the matrix
uses either linear or quadratic interpolation on meshes with
$h_c=0.2~\mathrm{m}$ and $h_f=0.1~\mathrm{m}$. Exact translational mortar
coupling is used and the electric field is set to zero. The imposed loading
is simple shear,
\begin{equation}
    \mathbf F_{\mathrm{macro}}(\ell)
    =
    \mathbf I+\ell\gamma_\star\mathbf e_1\otimes\mathbf e_2,
    \qquad
    \gamma_\star=0.25,
    \qquad
    \mathbf E_{\mathrm{macro}}=\mathbf 0 ,
\end{equation}
and the volumetric penalty is varied over
$K_m/\mu_m\in\{10,100,1000\}$. For each case, the volumetric term is
evaluated with and without J-bar averaging. The mesh sensitivity is
measured from the complete macroscopic stress tensor as
\begin{equation}
    d_h=
    \frac{
    \|\overline{\boldsymbol{\sigma}}_c-
    \overline{\boldsymbol{\sigma}}_f\|_F
    }{
    \|\overline{\boldsymbol{\sigma}}_f\|_F
    }.
\end{equation}

\begin{figure}[tbp]
    \centering
    \includegraphics[width=\linewidth]
    {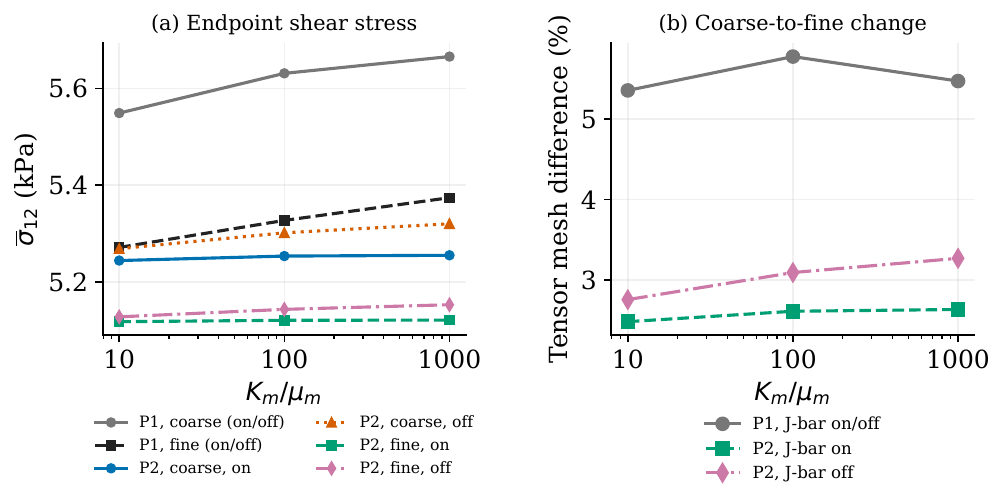}
    \caption{Sensitivity of the curved-cross response to matrix
    interpolation order, matrix resolution, and volumetric averaging.
    The left panel shows the endpoint shear stress and the right panel
    the relative coarse-to-fine change of the complete macroscopic
    stress tensor.}
    \label{fig:jbar-sensitivity}
\end{figure}

Figure~\ref{fig:jbar-sensitivity} shows a consistently smaller
coarse-to-fine change for quadratic interpolation over the tested
bulk-modulus range. With J-bar averaging, $d_h$ ranges from $2.48\%$ to
$2.63\%$, compared with $5.36\%$--$5.78\%$ for the linear
discretization. Removing J-bar averaging gives quadratic
coarse-to-fine differences of $2.76\%$--$3.27\%$. The linear results with
and without averaging coincide, as expected from their elementwise
constant deformation gradient. Based on this comparison, quadratic
interpolation with J-bar averaging is retained for the subsequent coupled
calculations. The complete factorial results are reported in
Appendix~\ref{app:verification-data}.

\subsection{Matrix Electric-Field Sampling Sensitivity}
\label{sec:matrix-field-sensitivity}

%
%
%
The mixed-dimensional electrical coupling evaluates the matrix electric
field along the beam centerlines. Unlike the mechanical mortar projection,
this operation samples the gradient of the discrete matrix potential at
one-dimensional locations, so its accuracy depends on the matrix
discretization and can also depend on where the beam lies relative to the
matrix elements. We examine these two effects separately. First, the
matrix mesh is refined while the physical network and loading are held
fixed. We then shift the matrix mesh relative to the same fixed physical
network without changing its geometry or connectivity. Together, these tests
measure the sensitivity of the homogenized response to the finite-resolution
sampling used in the matrix-to-beam electric coupling.

\subsubsection{Matrix-Mesh Resolution}
\label{sec:matrix-mesh-resolution}

%
%
%
%
%
%
%

The single-fiber electrical sweep in
Sec.~\ref{sec:single-fiber-discretization} samples an almost uniform matrix
electric field and therefore does not fully exercise the field projection in
a heterogeneous state. We examine this next using the curved cross under
combined mechanical and electrical loading. The cell is subjected to
$\mathbf F_{\mathrm{macro}}
=\mathbf I+0.15\,\mathbf e_1\otimes\mathbf e_2$ and a fixed transverse bias
$\mathbf E_{\mathrm{macro}}=25\,\mathbf e_3~\mathrm{MV/m}$, which produces
a spatially varying matrix electric field around the curved network. The
beam geometry, radius, beam discretization, material parameters, coupling
treatment, and loading are kept fixed while only the matrix mesh is refined.

The sensitivity of the electrical response is measured relative to the
finest tested matrix mesh as
\begin{equation}
e_D(h)
=
\frac{
\left\|
\overline{\mathbf D}_h
-
\overline{\mathbf D}_{h_f}
\right\|_2
}{
\left\|
\overline{\mathbf D}_{h_f}
\right\|_2
}.
\label{eq:matrix-mesh-sensitivity}
\end{equation}
For $h=0.20~\mathrm{m}$ and $0.15~\mathrm{m}$, the homogenized electric
displacement differs from the $h_f=0.10~\mathrm{m}$ result by $0.571\%$
and $0.415\%$, respectively. The matrix-to-beam electrical response is
therefore only weakly affected by matrix refinement over the tested range.
The differences are measured relative to the finest tested mesh,
$h_f=0.10~\mathrm{m}$. The complete endpoint quantities are reported in
Appendix~\ref{app:verification-data}.

\subsubsection{Relative Beam--Matrix-Mesh Placement}
\label{sec:placement-sensitivity}

%
%
%
%
%
%
%

Matrix refinement tests the resolution of the sampled electric field, but
not its dependence on the position of the beam centerline relative to the
matrix-element boundaries. This is relevant because the projected field at
a beam quadrature point is evaluated from a single containing matrix
element, without averaging gradients across neighboring elements. We
therefore keep the physical periodic straight-cross network fixed and
translate the matrix mesh relative to the network. All physical and
numerical settings are held fixed, so only the location of the beam
quadrature points within the matrix mesh changes.
Across the tested relative offsets, the homogenized stress tensor changes by
at most $0.069\%$ and the homogenized electric-displacement vector by
$0.120\%$ relative to the unshifted configuration. The response is therefore
only weakly dependent on the relative beam--matrix-mesh placement for this
mesh and loading. Together, the refinement and placement studies show only weak sensitivity
to matrix resolution and relative beam--matrix-mesh placement for the cases
considered here. The relative offsets and complete homogenized stress and
electric-displacement values are reported in
Appendix~\ref{app:verification-data}.

\section{Architecture-Dependent Electromechanical Response}
\label{sec:architecture-response}

%
%
%

Having assessed the numerical choices underlying the mixed-dimensional
formulation in Sec.~\ref{sec:numerical-verification}, we next examine how
the explicit fiber-network geometry influences the effective
electromechanical response. Unless stated otherwise, the calculations use
the common material parameters and numerical baseline introduced in
Sec.~\ref{sec:baseline-coefficients}. Mechanical behavior is characterized
primarily through the homogenized RVE stress, while the electrical response
is evaluated from deformation-induced changes in the homogenized
referential electric displacement under a fixed macroscopic electric bias.
Because the networks differ in orientation, curvature, connectivity, and
fiber content, the comparisons reflect the combined effect of these
architectural differences.
We begin with a network of straight aligned fibers, for which the relation
between fiber direction and macroscopic loading is particularly transparent.
Curved and multidirectional structured networks are then used to examine
how more complex centerline geometry changes the mechanical and electrical
response. Finally, the formulation is applied to irregular Voronoi-based
networks that retain the connectivity and orientational complexity of
nonwoven fiber systems.

\subsection{Aligned Fiber Network}
\label{sec:aligned-network}

\begin{figure}[!htbp]
    \centering
    \includegraphics[width=0.48\linewidth]
    {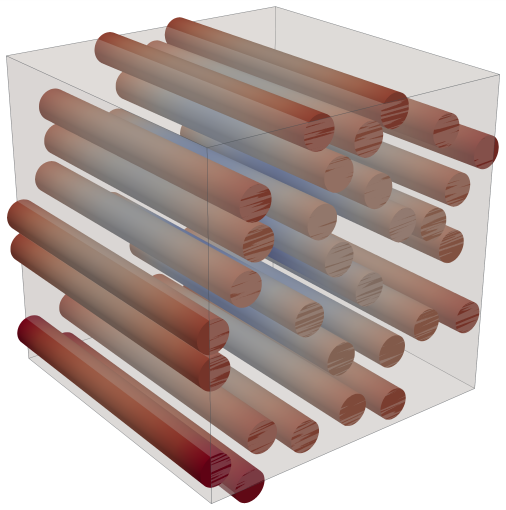}
    \caption{Undeformed aligned-fiber network containing 30 $x$-directed
    fibers with radius $r=0.05$ and total nominal geometric fiber volume
    fraction 23.56\%.}
    \label{fig:transverse_fibers_geometry}
\end{figure}

\begin{figure}[!htbp]
    \centering
    \begin{subfigure}[t]{0.48\linewidth}
        \centering
        \begin{pairedfieldpanel}
        \pairedfieldviews{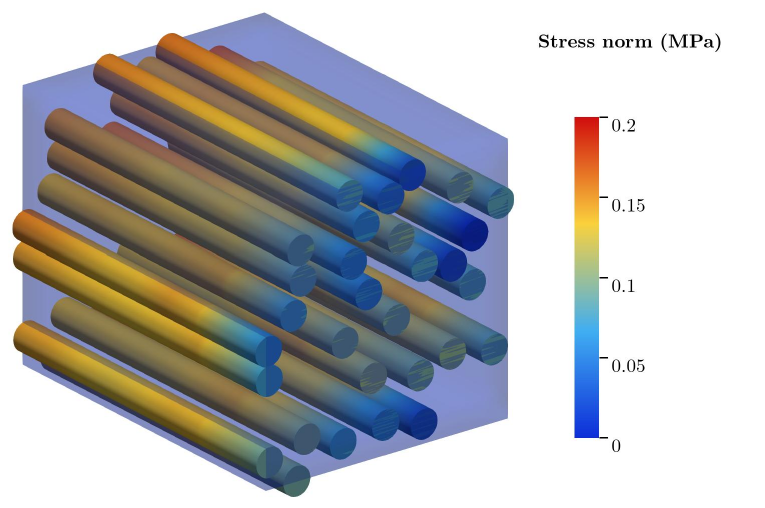}
        \end{pairedfieldpanel}
        \caption{$x$-extension at the isochoric target.}
        \label{fig:transverse_fibers_x_displacement}
    \end{subfigure}\hfill
    \begin{subfigure}[t]{0.48\linewidth}
        \centering
        \begin{pairedfieldpanel}
        \pairedfieldviews{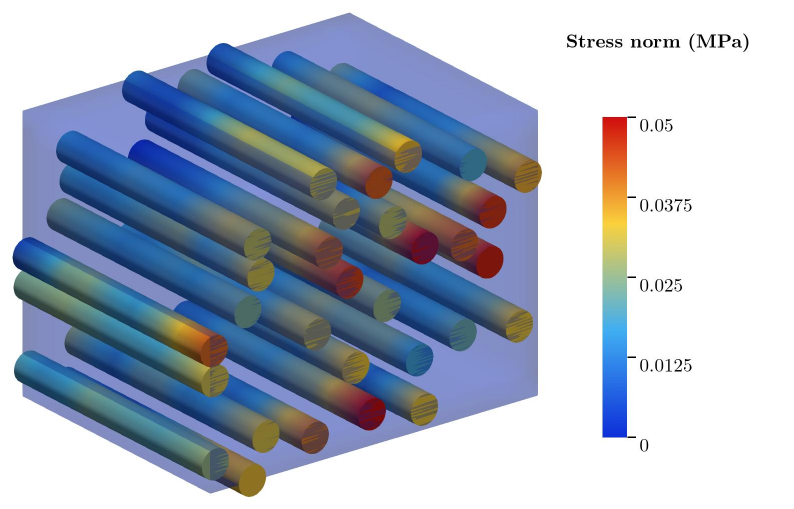}
        \end{pairedfieldpanel}
        \caption{$y$-extension at the isochoric target.}
        \label{fig:transverse_fibers_y_displacement}
    \end{subfigure}
    \caption{Final deformed configurations for the 30-fiber aligned network
    at matched $x$- and $y$-directed isochoric uniaxial targets. Each panel
    pairs total displacement (above) with local stress norm (below).}
    \label{fig:transverse_fibers_xy_displacement}
\end{figure}

The aligned-fiber RVE provides a direct comparison of mechanical and
electrical response parallel and transverse to the fiber direction. The
network contains 30 nonoverlapping $x$-directed fibers with radius
$r=0.05~\mathrm{m}$ and a nominal geometric fiber volume fraction of
approximately $23.56\%$, as shown in 
Fig.~\ref{fig:transverse_fibers_geometry}. To examine the directional
mechanical response, we apply matched isochoric extension targets parallel
and transverse to the fibers,
\begin{equation}
\mathbf{F}_{\star}^{(x)} =
\begin{bmatrix}
F_{11} & 0 & 0 \\
0 & F_{11}^{-1/2} & 0 \\
0 & 0 & F_{11}^{-1/2}
\end{bmatrix},
\qquad F_{11}=1.1,
\label{eq:transverse-fiber-x-loading}
\end{equation}
and
\begin{equation}
\mathbf{F}_{\star}^{(y)} =
\begin{bmatrix}
F_{22}^{-1/2} & 0 & 0 \\
0 & F_{22} & 0 \\
0 & 0 & F_{22}^{-1/2}
\end{bmatrix},
\qquad F_{22}=1.1.
\label{eq:transverse-fiber-y-loading}
\end{equation}
The loaded normal stress is
$\overline{\sigma}_{11,\mathrm{RVE}}=27.72~\mathrm{kPa}$ for
$x$-extension and
$\overline{\sigma}_{22,\mathrm{RVE}}=4.012~\mathrm{kPa}$ for
$y$-extension. When the load is applied along the fibers, they directly
carry the imposed stretch. Under transverse loading, the loaded direction
is primarily matrix dominated, while the prescribed lateral contraction
compresses the fibers along $x$. The corresponding final configurations
are shown in Fig.~\ref{fig:transverse_fibers_xy_displacement}, and the
complete macroscopic stress tensors are reported in
Table~\ref{tab:transverse_stress_tensors}.

\begin{figure}[!htbp]
    \centering
    \begin{subfigure}[t]{0.48\linewidth}
        \centering
        \includegraphics[width=\linewidth]
        {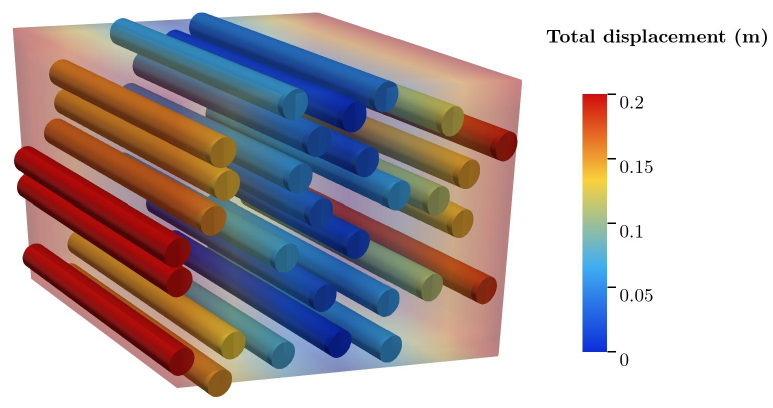}
        \caption{Total displacement.}
    \end{subfigure}\hfill
    \begin{subfigure}[t]{0.48\linewidth}
        \centering
        \includegraphics[width=\linewidth]
        {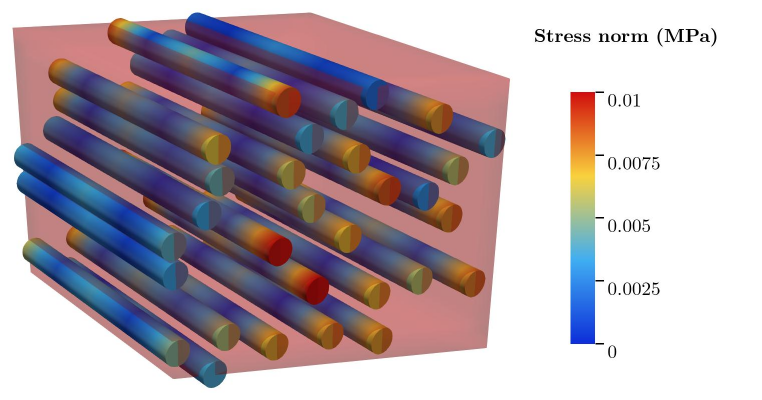}
        \caption{Local stress norm.}
    \end{subfigure}
    \caption{Final deformed 30-fiber aligned network under simple shear
    with final $F_{12}=0.25$.}
    \label{fig:transverse_fibers_shear_displacement}
\end{figure}

The role of the fiber direction is also evident under simple shear,
\begin{equation}
\mathbf{F}_{\mathrm{macro}}^{(s)} =
\begin{bmatrix}
1 & F_{12} & 0 \\
0 & 1 & 0 \\
0 & 0 & 1
\end{bmatrix},
\qquad 0 \le F_{12} \le 0.25.
\label{eq:transverse-fiber-shear-loading}
\end{equation}
For this loading, $\mathbf F_{\mathrm{macro}}\mathbf e_1=\mathbf e_1$,
so the affine $x$-directed fibers do not stretch. At $F_{12}=0.25$, the
homogenized shear stress is
$\overline{\sigma}_{12,\mathrm{RVE}}=5.000~\mathrm{kPa}$ and agrees with
the matrix-only response at the reported precision. The contrast with the
extension cases shows that the contribution of the aligned fibers depends
on how the imposed deformation acts relative to their orientation rather
than simply on their presence in the RVE. The final shear deformation is
shown in Fig.~\ref{fig:transverse_fibers_shear_displacement}.

\begin{figure}[!htbp]
    \centering
    \includegraphics[width=0.86\linewidth]
    {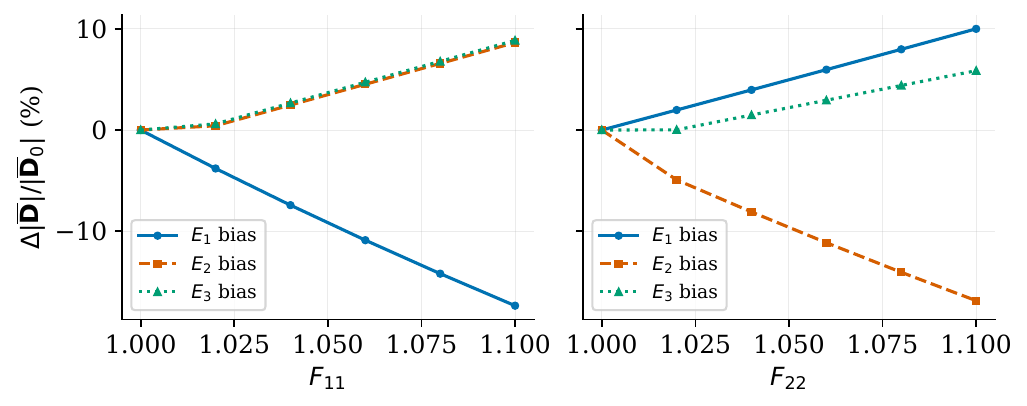}
    \caption{Fixed-bias sensor response for the 30-fiber aligned
    network. The two panels compare $x$-tension and $y$-tension toward
    isochoric targets; within each panel the electric bias is fixed along
    $E_1$, $E_2$, or $E_3$. The response is plotted as the relative change
    in $|\overline{\mathbf{D}}_{\mathrm{RVE}}|$ from the initially biased
    state.}
    \label{fig:transverse_fibers_sensor_bias_matrix}
\end{figure}

The same directional dependence appears in the fixed-bias
electrical response. The two extension paths above are each combined with
biases along the three coordinate directions,
\begin{equation}
\mathbf{E}_{\mathrm{target}} \in
\left\{
(25,0,0),\; (0,25,0),\; (0,0,25)
\right\}~\mathrm{MV/m}.
\label{eq:transverse-fiber-sensor-biases}
\end{equation}
This gives six combinations of mechanical loading and electric-bias
direction. Figure~\ref{fig:transverse_fibers_sensor_bias_matrix} shows the
relative change in the direction-invariant homogenized
electric-displacement magnitude,
$|\overline{\mathbf D}_{\mathrm{RVE}}|$, from the initially biased state.
Under $x$-extension, the $E_1$ bias separates from the two transverse bias
directions, while under $y$-extension both the sign and magnitude of the
response change. The aligned network therefore produces directional
mechanical and electrical responses, with the fixed-bias electrical signal
depending on both the deformation direction and the electric-bias
direction.

\subsection{Curved and Multidirectional Structured Networks}
\label{sec:structured-networks}

\begin{figure}[!htbp]
    \centering
    \begin{subfigure}[t]{0.31\linewidth}
        \centering
        \begin{alignedpanel}
        \includegraphics[width=\linewidth,height=0.85\linewidth,keepaspectratio]
        {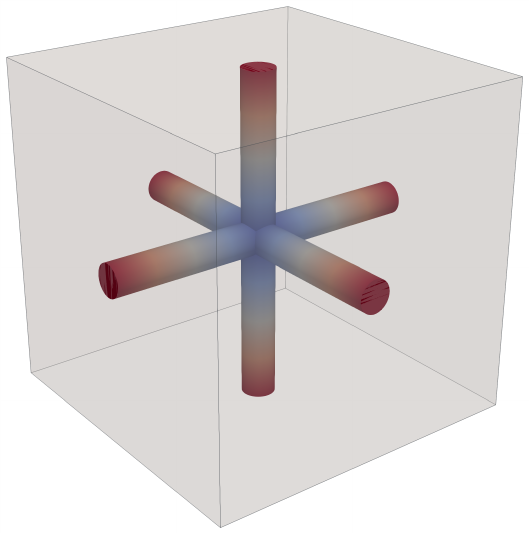}
        \end{alignedpanel}
        \caption{Straight center cross.}
        \label{fig:lattice_straight_base_mesh}
    \end{subfigure}\hfill
    \begin{subfigure}[t]{0.31\linewidth}
        \centering
        \begin{alignedpanel}
        \includegraphics[width=\linewidth,height=0.85\linewidth,keepaspectratio]
        {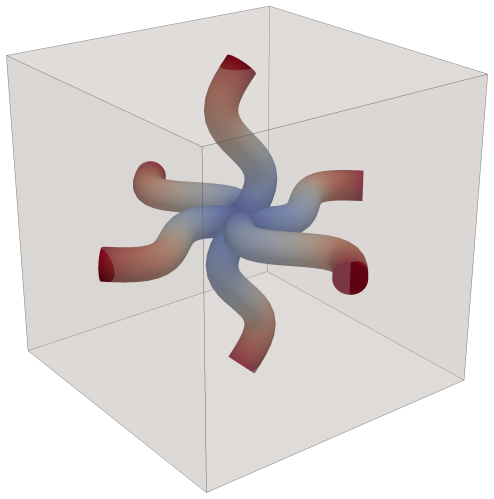}
        \end{alignedpanel}
        \caption{Curved center cross.}
        \label{fig:lattice_curved_base_mesh}
    \end{subfigure}\hfill
    \begin{subfigure}[t]{0.31\linewidth}
        \centering
        \begin{alignedpanel}
        \includegraphics[width=\linewidth,height=0.85\linewidth,keepaspectratio]
        {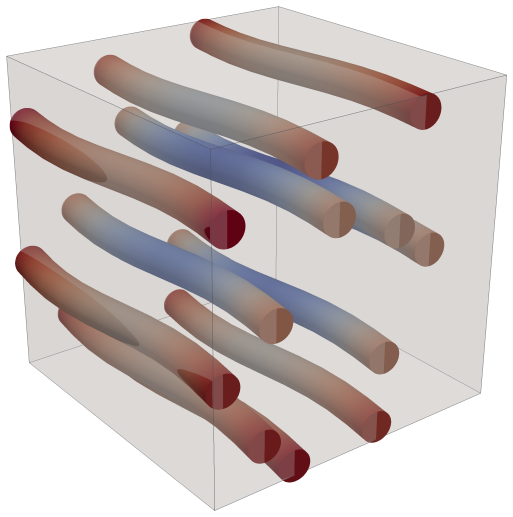}
        \end{alignedpanel}
        \caption{Wavy aligned network.}
        \label{fig:weave_base_mesh}
    \end{subfigure}
    \caption{Undeformed straight center cross, curved center cross, and
    wavy aligned network used for the structured-network comparisons.}
    \label{fig:lattice_geometry_montage}
\end{figure}

\begin{figure}[!htbp]
    \centering
    \begin{subfigure}[t]{0.31\linewidth}
        \centering
        \begin{pairedfieldpanel}
        \pairedfieldviews{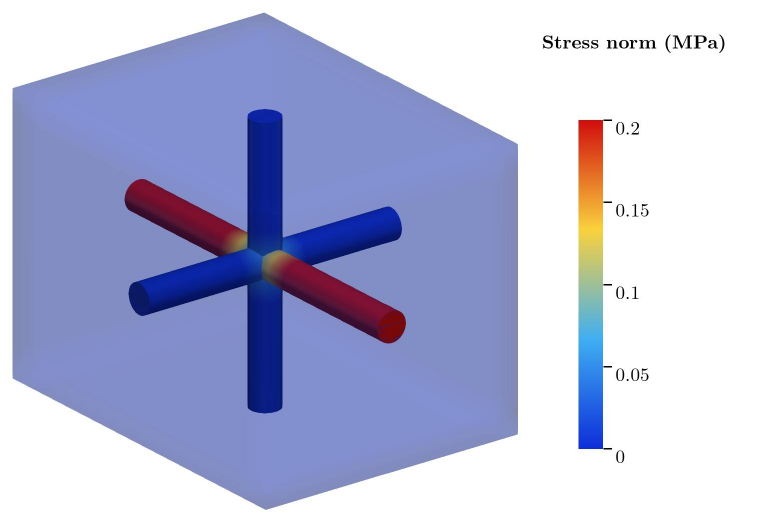}
        \end{pairedfieldpanel}
        \caption{Straight center cross.}
        \label{fig:lattice_straight_uniaxial_deformation}
    \end{subfigure}\hfill
    \begin{subfigure}[t]{0.31\linewidth}
        \centering
        \begin{pairedfieldpanel}
        \pairedfieldviews{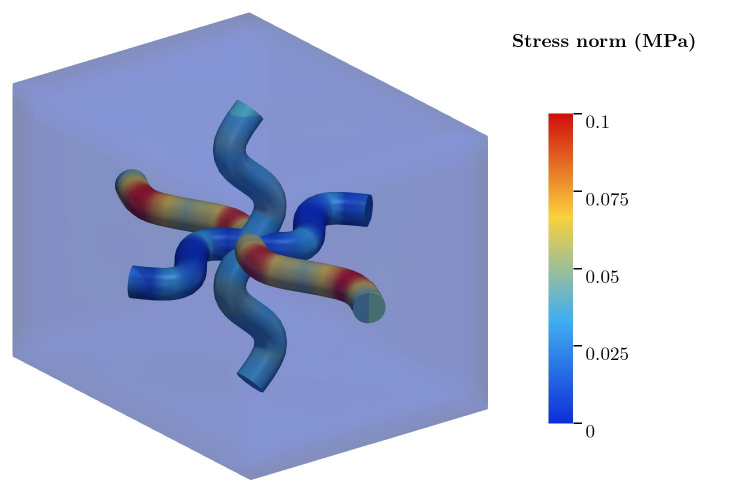}
        \end{pairedfieldpanel}
        \caption{Curved center cross.}
        \label{fig:lattice_curved_uniaxial_deformation}
    \end{subfigure}\hfill
    \begin{subfigure}[t]{0.31\linewidth}
        \centering
        \begin{pairedfieldpanel}
        \pairedfieldviews{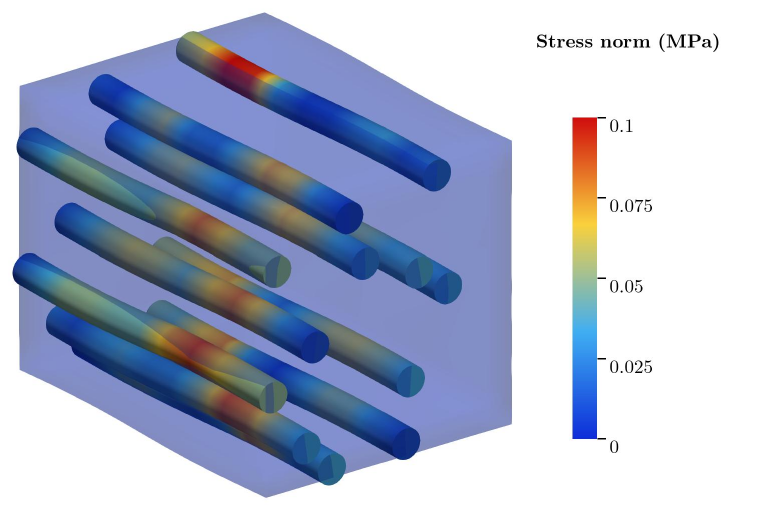}
        \end{pairedfieldpanel}
        \caption{Wavy aligned network.}
        \label{fig:weave_uniaxial_deformation}
    \end{subfigure}
    \caption{Final deformed structured networks at the common isochoric
    $x$-extension target with $F_{11}=1.1$. Each panel pairs total
    displacement (above) with local stress norm (below).}
    \label{fig:lattice_uniaxial_deformation}
\end{figure}

The aligned network in Sec.~\ref{sec:aligned-network} has a single fiber
direction, so its directional response can be related directly to the
orientation of the fibers. We next consider networks in which this relationship is less
direct because the fibers are distributed over several directions or have
curved centerlines. The three cases are a straight three-axis center cross,
a curved center cross with the same global connectivity, and a wavy aligned
network composed of parallel sinusoidal paths, as shown in
Fig.~\ref{fig:lattice_geometry_montage}. The straight and curved center
crosses provide the closest comparison because both contain three
face-to-face beam families meeting at a common center vertex, while the
curved case replaces the straight centerlines by sinusoidal paths. The wavy
aligned network retains an aligned arrangement but also introduces
sinusoidal centerlines over a larger fiber population. Their nominal geometric fiber volume
fractions are approximately $2.36\%$, $2.81\%$, and $9.48\%$ for the
straight cross, curved cross, and wavy aligned network, respectively. To compare how the three networks carry the same
macroscopic deformation, we first apply the common isochoric $x$-extension
target
\begin{equation}
\mathbf{F}_{\star} =
\begin{bmatrix}
F_{11} & 0 & 0 \\
0 & F_{11}^{-1/2} & 0 \\
0 & 0 & F_{11}^{-1/2}
\end{bmatrix},
\qquad F_{11}=1.1 .
\label{eq:lattice-weave-uniaxial-load}
\end{equation}
The loaded normal stress
$\overline{\sigma}_{11,\mathrm{RVE}}$ is
$4.803~\mathrm{kPa}$ for the straight cross,
$4.358~\mathrm{kPa}$ for the curved cross, and
$13.14~\mathrm{kPa}$ for the wavy aligned network. The two center-cross
architectures therefore give similar axial responses, whereas the wavy
aligned network carries a substantially larger load in the $x$ direction.
The wavy aligned network also differs in fiber amount and organization, so
the larger response reflects the complete architecture. The corresponding
final configurations are shown in
Fig.~\ref{fig:lattice_uniaxial_deformation}, and the complete macroscopic
stress tensors are reported in Table~\ref{tab:lattice_stress_tensors}.

\begin{figure}[!htbp]
    \centering
    \begin{subfigure}[t]{0.31\linewidth}
        \centering
        \begin{pairedfieldpanel}
        \pairedfieldviews{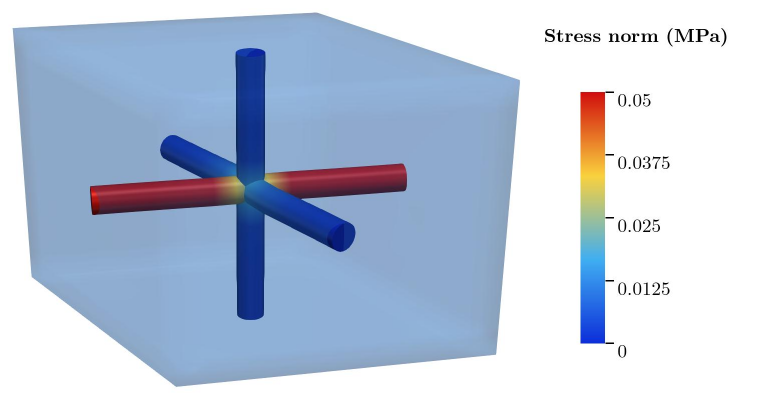}
        \end{pairedfieldpanel}
        \caption{Straight center cross.}
        \label{fig:lattice_straight_shear_deformation}
    \end{subfigure}\hfill
    \begin{subfigure}[t]{0.31\linewidth}
        \centering
        \begin{pairedfieldpanel}
        \pairedfieldviews{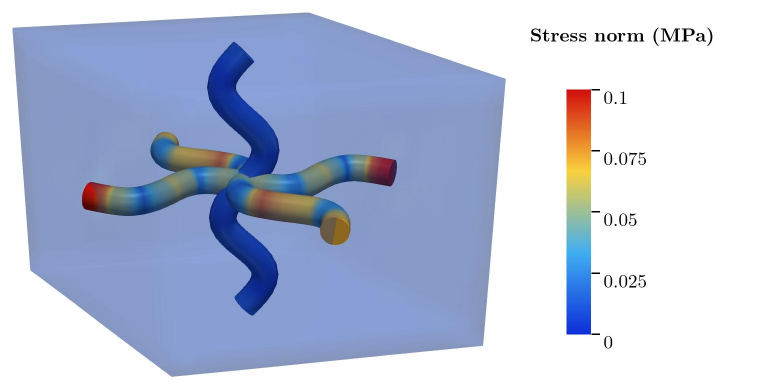}
        \end{pairedfieldpanel}
        \caption{Curved center cross.}
        \label{fig:lattice_curved_shear_deformation}
    \end{subfigure}\hfill
    \begin{subfigure}[t]{0.31\linewidth}
        \centering
        \begin{pairedfieldpanel}
        \pairedfieldviews{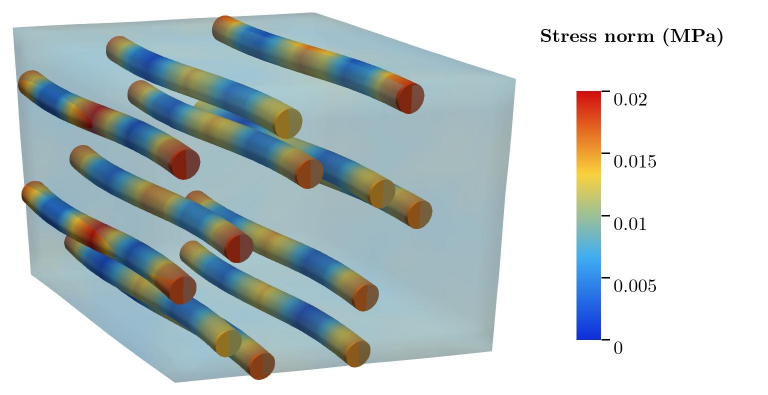}
        \end{pairedfieldpanel}
        \caption{Wavy aligned network.}
        \label{fig:weave_shear_deformation}
    \end{subfigure}
    \caption{Final deformed structured networks under simple shear with
    $F_{12}=0.25$. Each panel pairs total displacement (above) with local
    stress norm (below).}
    \label{fig:lattice_shear_deformation}
\end{figure}

Simple shear provides a second mechanical loading mode that is distinct
from extension along the global $x$ direction and therefore allows the
different fiber directions and curved centerlines to contribute differently
to the response. We prescribe
\begin{equation}
\mathbf{F}_{\mathrm{macro}} =
\begin{bmatrix}
1 & F_{12} & 0 \\
0 & 1 & 0 \\
0 & 0 & 1
\end{bmatrix},
\qquad 0 \leq F_{12} \leq 0.25 .
\label{eq:lattice-weave-shear-load}
\end{equation}
At $F_{12}=0.25$, the homogenized shear stress
$\overline{\sigma}_{12,\mathrm{RVE}}$ is
$5.083~\mathrm{kPa}$ for the straight cross,
$5.254~\mathrm{kPa}$ for the curved cross, and
$5.093~\mathrm{kPa}$ for the wavy aligned network. The three shear
responses are therefore of similar magnitude despite the substantially
different axial responses under $x$-extension. Their normal stress
components and local deformation patterns nevertheless differ, as shown in
Fig.~\ref{fig:lattice_shear_deformation}. The architecture dependence therefore varies with loading mode: the
networks differ strongly in extension but much less in shear.

\begin{figure}[htbp]
    \centering
    \includegraphics[width=0.94\linewidth]
    {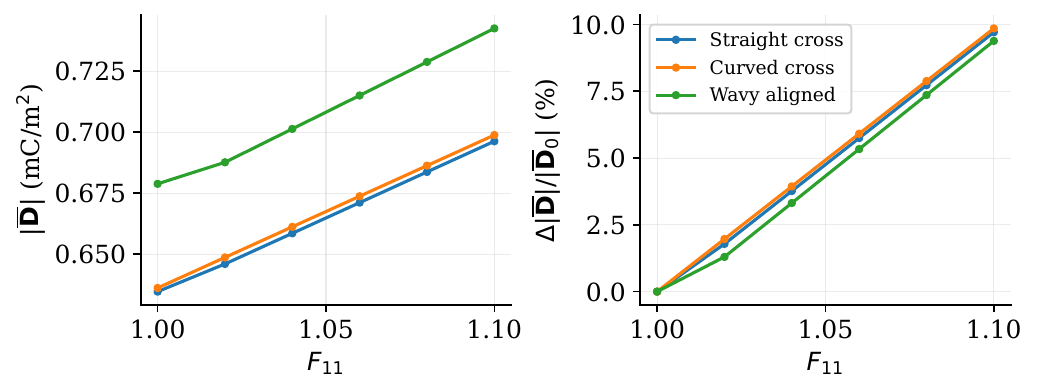}
    \caption{Electric-displacement magnitude and relative
    fixed-bias response of the straight cross, curved cross,
    and wavy aligned network under their shared mechanical loading path.}
    \label{fig:lattice-weave-electric}
\end{figure}

We finally compare the fixed-bias sensing response of the same
three networks under a common mechanical and electrical loading. All three networks are subjected to the same mechanical loading and electric
bias so that their electrical responses can be compared directly. The electric field is held fixed along $\mathbf e_3$ while the
cell is extended in the $x$ direction,
\begin{equation}
\mathbf E_{\mathrm{macro}}(t)=E_0\mathbf e_3,
\qquad
\mathbf F_{\mathrm{macro}}(t)
=(1-t)\mathbf I+t\mathbf F_{\mathrm{macro},\star},
\qquad t\in[0,1],
\end{equation}
with
\begin{equation}
\mathbf F_{\mathrm{macro},\star}
=
\operatorname{diag}
\left(
\lambda_\star,
\lambda_\star^{-1/2},
\lambda_\star^{-1/2}
\right),
\qquad
\lambda_\star=1.10,
\qquad
E_0=25~\mathrm{MV/m}.
\label{eq:lattice-weave-sensor-biases}
\end{equation}
Figure~\ref{fig:lattice-weave-electric} shows both the homogenized
electric-displacement magnitude and its change relative to the initially
biased state. The curved and straight center crosses show the largest
relative changes, respectively, while the wavy aligned network shows the
smallest relative change despite having the largest absolute
electric-displacement magnitude.

\subsection{Irregular Fiber Networks}
\label{sec:irregular-networks}

The preceding examples use prescribed fiber families whose response can be
related directly to a small number of geometric features such as alignment
and centerline curvature. We next consider irregular networks for which this
direct interpretation is no longer available. The Voronoi-based
architectures contain thousands of connected segments with distributed
orientations, branch lengths, and junctions, so their effective response
emerges from the collective network geometry rather than from a small
number of prescribed fiber directions. These comparisons therefore characterize the response of the complete
network architectures rather than the effect of a single geometric
parameter.

The irregular networks are motivated by electrospun and related nonwoven
fiber systems, where orientation distribution, connectivity, and local fiber
density can all influence the effective response. We consider a nominally
isotropic network, a network with a preferred in-plane fiber direction, and
a perforated network in which the fibers are embedded in a matrix containing
a central void. 
The irregular networks are
periodic Voronoi-based centerline graphs. Their construction, filtering,
junction treatment, and detailed topology metadata are described in
Appendix~\ref{app:voronoi-generation}.
They differ in several geometric descriptors,
and the perforated case additionally changes the matrix domain. 

\subsubsection{Network Architectures}
\label{sec:voronoi-architectures}

\begin{figure}[tbp]
    \centering
    \begin{subfigure}[t]{0.49\linewidth}
        \centering
        \begin{alignedpanel}
        \includegraphics[
            width=\linewidth,
            height=0.85\linewidth,
            keepaspectratio
        ]{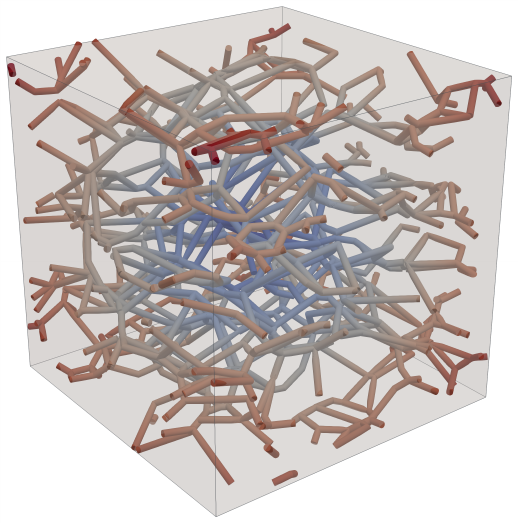}
        \end{alignedpanel}
        \caption{Oblique view.}
    \end{subfigure}\hfill
    \begin{subfigure}[t]{0.49\linewidth}
        \centering
        \begin{alignedpanel}
        \includegraphics[
            width=\linewidth,
            height=0.85\linewidth,
            keepaspectratio
        ]{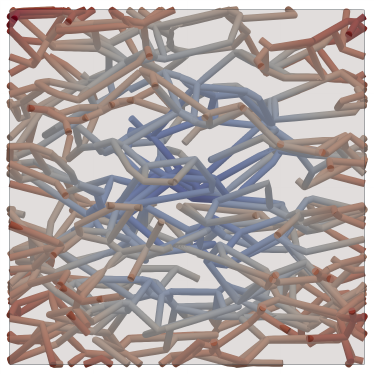}
        \end{alignedpanel}
        \caption{Side view.}
    \end{subfigure}
    \caption{Preferred-direction Voronoi network. The network construction
biases the fiber orientations toward the $x$--$y$ plane.}
    \label{fig:voronoi_preferred_geometry}
\end{figure}

We consider three irregular network architectures before comparing their
directional mechanical and fixed-bias electrical responses. The first is a
nominally isotropic finite random network generated without an imposed
orientation filter. It contains 4619 beam elements and has a nominal geometric fiber volume
fraction of $f_b^{\mathrm{nom}}=4.55\%$. This network provides a case without a
prescribed preferred direction, so the directional loading considered below
can be used to assess how closely the effective response of this particular
finite realization approaches direction independence. The second network introduces a preferred in-plane orientation by
selecting segments with unit tangent $\mathbf t$ that satisfy
\begin{equation}
    |t_z| \leq t_z^{\max},
    \qquad
    t_z^{\max}=0.7 .
    \label{eq:voronoi-preferred-direction-filter}
\end{equation}
The filter biases the retained segments toward the $x$--$y$ plane, producing
an irregular network with a prescribed in-plane orientation bias. The resulting network contains 3103 beam elements and has
$f_b^{\mathrm{nom}}=2.97\%$. The orientation filter also changes the network connectivity, total
centerline length, and fiber content. Its
geometry is shown in Fig.~\ref{fig:voronoi_preferred_geometry}.


\begin{figure}[tbp]
    \centering
    \begin{subfigure}[t]{0.49\linewidth}
        \centering
        \begin{alignedpanel}
        \includegraphics[
            width=\linewidth,
            height=0.85\linewidth,
            keepaspectratio
        ]{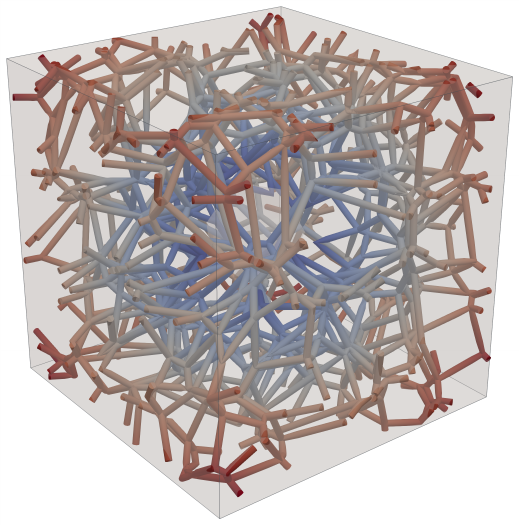}
        \end{alignedpanel}
        \caption{Oblique view.}
    \end{subfigure}\hfill
    \begin{subfigure}[t]{0.49\linewidth}
        \centering
        \begin{alignedpanel}
        \includegraphics[
            width=\linewidth,
            height=0.85\linewidth,
            keepaspectratio
        ]{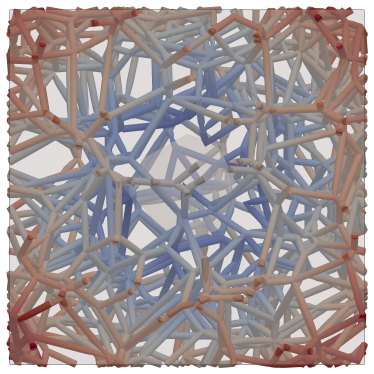}
        \end{alignedpanel}
        \caption{Side view.}
    \end{subfigure}
    \caption{Perforated Voronoi network with a central insulating void and
    the surrounding irregular fiber network.}
    \label{fig:voronoi_perforated_geometry}
\end{figure}

The third architecture combines an irregular network with a matrix
containing a central insulating void. The void has radius $0.15~\mathrm{m}$,
and the fiber centerlines are generated with a clearance of $0.01~\mathrm{m}$
from the excluded region. No mechanical or electrical field is solved
inside the void; the internal matrix--void surface carries the natural
traction-free and electrically insulating boundary conditions introduced
in Sec.~\ref{sec:theoretical-setting}. The resulting network contains 4766 beam elements and has
$f_b^{\mathrm{nom}}=4.61\%$, as shown in
Fig.~\ref{fig:voronoi_perforated_geometry}. The perforated case uses a separate network realization in addition to the
change in matrix geometry. Additional network metadata are reported in
Table~\ref{tab:network-metadata}.

\subsubsection{Directional Mechanical Response}
\label{sec:voronoi-mechanical}

\begin{figure}[tbp]
    \centering
    \begin{subfigure}[t]{0.48\linewidth}
        \centering
        \begin{pairedfieldpanel}
        \pairedfieldviews{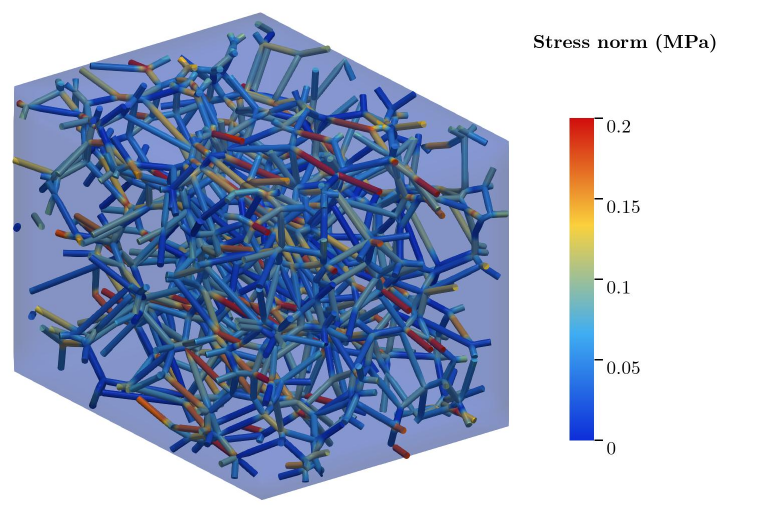}
        \end{pairedfieldpanel}
        \caption{$x$ extension.}
    \end{subfigure}\hfill
    \begin{subfigure}[t]{0.48\linewidth}
        \centering
        \begin{pairedfieldpanel}
        \pairedfieldviews{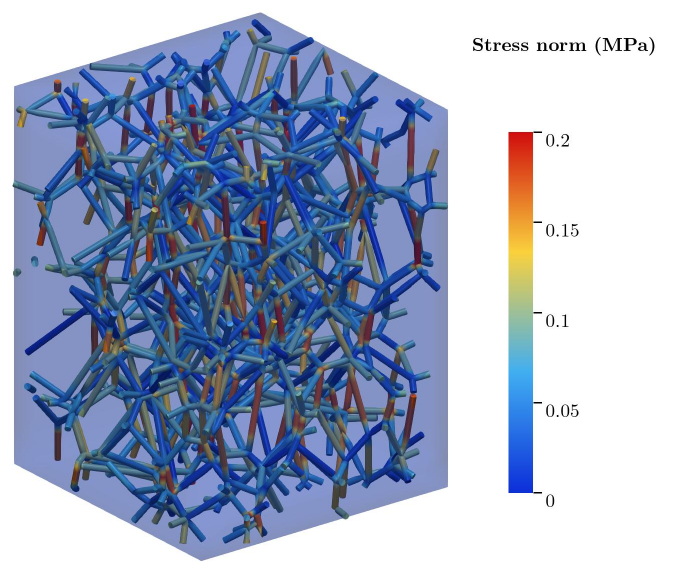}
        \end{pairedfieldpanel}
        \caption{$z$ extension.}
    \end{subfigure}
    \caption{Representative extension states of the nominally isotropic
    Voronoi network. Each panel pairs total displacement (above) with local
    stress norm (below).}
    \label{fig:voronoi_isotropic_deformation}
\end{figure}

\begin{figure}[tbp]
    \centering
    \begin{subfigure}[t]{0.48\linewidth}
        \centering
        \begin{pairedfieldpanel}
        \pairedfieldviews{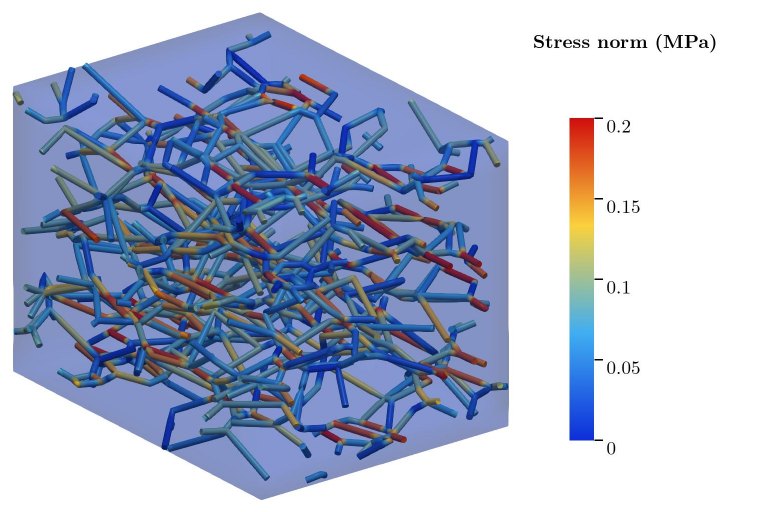}
        \end{pairedfieldpanel}
        \caption{$x$ extension.}
    \end{subfigure}\hfill
    \begin{subfigure}[t]{0.48\linewidth}
        \centering
        \begin{pairedfieldpanel}
        \pairedfieldviews{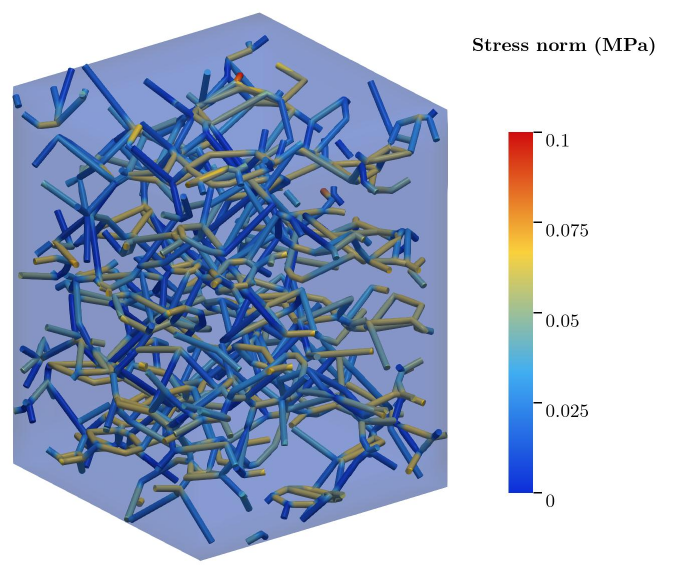}
        \end{pairedfieldpanel}
        \caption{$z$ extension.}
    \end{subfigure}
    \caption{Representative in-plane and out-of-plane extension states of
    the preferred-direction Voronoi network. Each panel pairs total
    displacement (above) with local stress norm (below).}
    \label{fig:voronoi_preferred_deformation}
\end{figure}

The structured networks in Sec.~\ref{sec:structured-networks} contain
prescribed fiber families whose relation to the loading direction can be
identified directly from their geometry. For the Voronoi networks, the
macroscopic response instead results from hundreds to thousands of
interconnected branches with distributed orientations. We therefore use
directional extension to examine whether the organization of these
irregular networks remains visible in the homogenized mechanical response.
The same isochoric extension target with a loaded stretch of $1.1$ is
applied along the coordinate directions considered for each network, and the corresponding
loaded normal component of the homogenized Cauchy stress is compared below.
The complete macroscopic stress tensors, together with the
additional shear cases, are reported in
Table~\ref{tab:voronoi_stress_tensors}.
For the nominally isotropic realization,
$\overline{\sigma}_{11,\mathrm{RVE}}=4.698~\mathrm{kPa}$ under
$x$-extension and
$\overline{\sigma}_{33,\mathrm{RVE}}=4.678~\mathrm{kPa}$ under
$z$-extension. The two loaded responses differ by less than $0.5\%$,
indicating little directional variation between the two sampled directions 
for this particular realization. The corresponding loaded configurations in
Fig.~\ref{fig:voronoi_isotropic_deformation} show how this response emerges
from the deformation of the irregular network rather than from a small
number of prescribed fiber paths.
The preferred-direction network shows a clearer directional separation.
Its loaded normal stresses are $4.699~\mathrm{kPa}$,
$4.702~\mathrm{kPa}$, and $3.999~\mathrm{kPa}$ under
$x$-, $y$-, and $z$-extension, respectively. The two in-plane responses are nearly identical, whereas the out-of-plane
response is approximately $15\%$ lower. This is consistent with the imposed
bias of the retained fiber segments toward the $x$--$y$ plane and shows that
the directional organization of an irregular connected network can remain
visible after homogenization. Figure~\ref{fig:voronoi_preferred_deformation}
shows representative in-plane and out-of-plane loaded states.

The perforated realization gives loaded normal stresses of
$4.718~\mathrm{kPa}$, $4.680~\mathrm{kPa}$, and $4.668~\mathrm{kPa}$
under $x$-, $y$-, and $z$-extension, respectively. The three responses
differ by less than $1.1\%$, so this particular realization exhibits little
directional variation over the sampled extension directions despite its
irregular local geometry and central void. Taken together, the three cases
show that irregular networks can range from nearly direction-independent
responses over the sampled directions to a clear in-plane/out-of-plane
distinction when a directional bias is introduced into the network
architecture.

\subsubsection{Fixed-Bias Electrical Response}
\label{sec:voronoi-electrical}

The directional mechanical results in Sec.~\ref{sec:voronoi-mechanical}
show that differences in the organization of the irregular networks can
remain visible after homogenization. We next examine whether the same is
true of the fixed-bias electrical response. Unlike the aligned
and structured examples, where the response can be related to a small
number of prescribed fiber paths, the electrical response here results from
the collective deformation of hundreds to thousands of connected branches.
To compare the three architectures under the same conditions, each network
is subjected to a fixed macroscopic electric bias
$\mathbf E_{\mathrm{macro}}=25\,\mathbf e_3~\mathrm{MV/m}$ while the
mechanical deformation is ramped in the $x$ direction toward the isochoric
target $F_{11}=1.1$. The electric field is present from the initially biased
state and remains fixed throughout the mechanical loading. The sensing
response is measured by the relative change in the homogenized
electric-displacement magnitude,
\begin{equation}
    \frac{\Delta
    |\overline{\mathbf D}_{\mathrm{RVE}}|}
    {|\overline{\mathbf D}_{\mathrm{RVE}}|_0}
    =
    \frac{
    |\overline{\mathbf D}_{\mathrm{RVE}}|
    -
    |\overline{\mathbf D}_{\mathrm{RVE}}|_0
    }{
    |\overline{\mathbf D}_{\mathrm{RVE}}|_0
    } ,
\end{equation}
where the subscript $0$ denotes the initially biased state. Each network is
normalized by its own initial value because the architectures differ in
fiber amount and, for the perforated case, in matrix domain. The resulting
quantity therefore expresses the deformation-induced electrical change
relative to the initial state of each architecture rather than comparing
their absolute electric-displacement magnitudes directly.

\begin{figure}[tbp]
    \centering
    \includegraphics[width=0.76\linewidth]
    {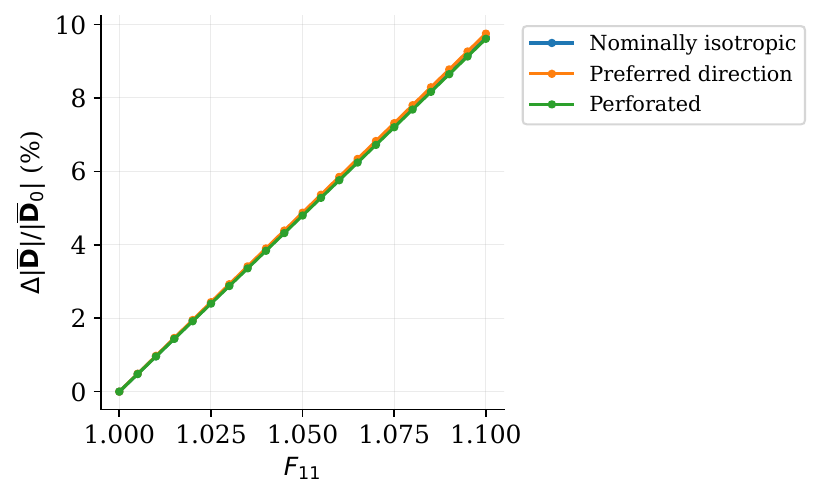}
    \caption{Fixed-bias response of the nominally isotropic,
    preferred-direction, and perforated Voronoi networks during
    $x$-extension toward the isochoric target $F_{11}=1.1$. The electric
    bias is fixed at
    $\mathbf E_{\mathrm{macro}}=25\,\mathbf e_3~\mathrm{MV/m}$, and the
    plotted response is the relative change in homogenized
    electric-displacement magnitude from the initially biased state.}
    \label{fig:voronoi_sensor_combined}
\end{figure}

Figure~\ref{fig:voronoi_sensor_combined} compares the complete sensing
curves for the nominally isotropic, preferred-direction, and perforated
networks. The three architectures produce slightly different responses under
the same macroscopic bias and deformation. At $F_{11}=1.1$, the relative
changes in
$|\overline{\mathbf D}_{\mathrm{RVE}}|$
are approximately $9.648\%$, $9.752\%$, and $9.611\%$ for the
nominally isotropic, preferred-direction, and perforated networks,
respectively. The preferred-direction network therefore develops the largest
change over this loading path, while the perforated network develops the
smallest. The three
architectures therefore give distinct homogenized electrical responses
under the same macroscopic loading. Numerical sensitivity of the sensing
observable to quadrature and penalty parameter is
reported in Table~\ref{tab:sensing-sensitivity}.

\section{Discussion}
\label{sec:discussion}

The numerical examples show that explicit network architecture influences
the homogenized electromechanical response. For aligned fibers, the
mechanical and electrical responses depend strongly on the relation between
fiber orientation and the applied loading. The structured-network examples
further show that differences between architectures depend on the loading
mode, with stronger separation under extension than under shear. In the
irregular networks, the in-plane-biased architecture exhibits a clear
distinction between in-plane and out-of-plane loading, while the other
realizations show comparatively little directional variation. The
fixed-bias electrical response also varies among the network architectures.
For the irregular networks, the normalized changes are approximately
$9.6$--$9.8\%$. The homogeneous ideal matrix gives a $10\%$ change under
the same fixed-bias isochoric loading, so the network responses represent
architecture-dependent deviations from this kinematic baseline. A
device-level assessment would additionally require a specific electrode
configuration, measurement circuit, and noise model.

The formulation involves several modeling approximations. First, the beam energy is superposed on the matrix rather than replacing a
finite matrix volume. This choice avoids constructing a fiber-shaped
exclusion region in the matrix mesh, which would become increasingly
cumbersome for curved, intersecting, and densely connected networks, but it
makes the model an additive embedded reinforcement rather than a
phase-partitioned composite. Equal beam and matrix properties therefore do
not recover a homogeneous material. For the straight-fiber loading considered here, the mixed-dimensional and
body-fitted models give similar macroscopic stresses. The two formulations differ in how the fiber volume is represented: the
embedded model superposes the beam energy on the matrix, whereas the
body-fitted model replaces the matrix within the fiber volume. A second
approximation arises from representing each finite-radius fiber by
one-dimensional beam kinematics. This approximation is best suited to
branches whose length and radius of curvature are large relative to the
fiber diameter. As shown by the geometric measures in
Table~\ref{tab:network-metadata}, some Voronoi branches are locally very
short or strongly curved relative to the fiber diameter. The local
three-dimensional fields near these branches may therefore be represented
less accurately. This is less critical for the present study, which focuses
primarily on network-scale and homogenized response quantities.

A related approximation enters the electrical coupling. The fibers do not
carry independent electric-potential degrees of freedom. The matrix electric field
is sampled along the beam centerline and enters the beam dielectric enthalpy,
which preserves two-way electromechanical coupling without adding beam
electric-potential degrees of freedom. This choice is consistent with the
reduced beam description, but it does not resolve finite-radius
electric-field variation around an individual fiber or an explicit
dielectric matrix--fiber interface. In addition, the gradient of a general
$H^1$ matrix potential does not possess a guaranteed trace on a
one-dimensional set in three dimensions. However, the matrix-mesh refinement and
relative-placement studies in Sec.~\ref{sec:matrix-field-sensitivity} show
only weak sensitivity over the tested range. A finite-radius or regularized projection would provide a more local
description of the electric field around the fiber. 
The present comparisons show architecture-dependent behavior across
networks that differ simultaneously in orientation, curvature,
connectivity, and fiber content. Isolating the contribution of individual
descriptors will require matched architecture families in which these
quantities are varied independently.
The formulation targets problems in which explicit network geometry is
important but fully resolved three-dimensional fiber meshes are
impractical. If the goal is to reproduce the response of a phase-partitioned composite,
an excess-energy or phase-replacement treatment could remove the overlap
between matrix and fiber energies. Richer section models and finite-radius electrical projections could allow
the formulation to handle more strongly curved, shorter, or otherwise
challenging fiber geometries.

\section{Conclusion}
\label{sec:conclusion}

We developed a mixed-dimensional finite-element formulation for
electroactive fiber networks embedded in soft dielectric matrices. The
fibers are represented by geometrically exact beams that are mechanically
coupled to the surrounding matrix through projected mortar constraints,
while their dielectric response is driven directly by the matrix electric
field sampled along the beam centerlines. This construction provides
two-way electromechanical coupling without introducing independent electric
potential degrees of freedom on the fibers and allows explicit network
geometry to be retained without constructing body-fitted three-dimensional
fiber meshes. In the present work, the formulation was embedded in a
periodic homogenization framework to compute effective stress and electric
displacement.
The numerical examples show that this explicit representation preserves
architecture-dependent electromechanical behavior across increasingly
complex networks. Aligned fibers produce strongly directional responses,
while curved, multidirectional, and irregular networks retain distinct
mechanical and fixed-bias electrical behavior that emerges from their
collective geometry. The formulation therefore provides a computational
framework for studying how orientation, curvature, connectivity, and
network topology influence the effective response of electroactive soft
composites, with the longer-term objective of treating fiber architecture
as a controllable design variable.

\section*{Acknowledgment}
This material is based upon work partially supported by the U.S. National Science Foundation under award No. 2452029 (JNF) and awards No. 2438943, 2235856, 2127925, 2046148 (MKR).
The opinions, findings, and conclusions, or recommendations expressed are those of the authors and do not necessarily reflect the views of the NSF.

The authors acknowledge the Texas Advanced Computing Center (TACC) at The University of Texas at Austin for providing computational resources that have contributed to the research results reported within this paper. URL: http://www.tacc.utexas.edu

\bibliographystyle{unsrt}
\bibliography{references}

\clearpage
\appendix

\section{Model Settings and Discrete Diagnostics}
\label{app:model-settings}

This appendix collects the common material parameters, numerical settings,
postprocessing conventions, and discrete diagnostics used throughout the
reported calculations.

\begin{table}[H]
\centering
\caption{Baseline material parameters used in the numerical examples. The
beam section is kinematically incompressible, while the matrix uses the
finite volumetric penalty introduced in Sec.~\ref{sec:matrix-electromechanical-model}.}
\label{tab:latest-run-parameters}
\begin{adjustbox}{max width=\tablewidth,center}
\begin{tabular}{@{}p{0.44\linewidth}p{0.18\linewidth}p{0.25\linewidth}@{}}
\toprule
Quantity & Symbol & Value \\
\midrule
\multicolumn{3}{@{}l}{\textit{Beam dielectric elastomer}} \\
Young's modulus & $Y_b$ & $1.0~\mathrm{MPa}$ \\
Modulus-conversion coefficient & $\nu_b^{\mathrm{input}}$ & $0.495$ \\
Shear modulus & $\mu_b$ & $0.334448~\mathrm{MPa}$ \\
Relative permittivity & $\varepsilon_r$ & $2.8$ \\
Beam dielectric coefficient & $\epsilon_e$
& $2.48\times10^{-5}~\mathrm{MPa}/(\mathrm{MV/m})^2$ \\
\addlinespace[2pt]
\multicolumn{3}{@{}l}{\textit{Matrix elastomer}} \\
Young's modulus & $Y_m$ & $0.060~\mathrm{MPa}$ \\
Shear modulus & $\mu_m$ & $0.0200~\mathrm{MPa}$ \\
Isochoric coefficient & $C_{1m}$ & $0.0100~\mathrm{MPa}$ \\
Volumetric penalty coefficient & $K_m$ & $2.0~\mathrm{MPa}$ \\
Relative permittivity & $\varepsilon_{r,m}$ & $2.8$ \\
Matrix dielectric coefficient & $\epsilon_m$
& $2.48\times10^{-5}~\mathrm{MPa}/(\mathrm{MV/m})^2$ \\
Beam-to-matrix stiffness ratio & $Y_b/Y_m$ & $16.7$ \\
\bottomrule
\end{tabular}
\end{adjustbox}
\end{table}

The common numerical settings are summarized in
Table~\ref{tab:numerical-settings}. Exact translational mortar constraints
are used for the smaller structured problems and verification studies,
whereas the large Voronoi networks use penalty coupling. In exact-mortar
calculations, the penalty parameter is used only during continuation and
does not enter the final constrained system.

\begin{table}[H]
\centering
\caption{Numerical settings used unless varied explicitly in the
verification studies.}
\label{tab:numerical-settings}
\begin{adjustbox}{max width=\tablewidth,center}
\begin{tabular}{@{}ll@{}}
\toprule
Quantity & Setting \\
\midrule
Matrix displacement and potential & Quadratic tetrahedra, linear geometry \\
Matrix volumetric treatment & Element-wise J-bar averaging \\
Beam translations and rotation vectors & Linear centerline elements \\
Mortar interpolation & Linear \\
Matrix quadrature & Simplex Gauss rule, order four \\
Embedded beam--matrix quadrature & Eight-point Gauss per matrix subsegment \\
Circular-section integration & Three radial Gauss--Legendre, twelve angular samples \\
Linear solver & Sparse direct UMFPACK \\
Nonlinear relative / absolute controls & $10^{-8}$ / $10^{-10}$ \\
Voronoi translational penalty & $p=10^6~\mathrm{MPa}$ \\
Penalty to matrix shear-modulus ratio & $p/\mu_m=5\times10^7$ \\
\bottomrule
\end{tabular}
\end{adjustbox}
\end{table}

The deformed field visualizations use the display mapping
\begin{equation}
    \mathbf{x}_{\mathrm{vis}}
    =
    \mathbf{X}+2\mathbf{u}(\mathbf{X}).
    \label{eq:visual-warp}
\end{equation}
The factor of two changes only the displayed geometry and does not scale the
field values. The local stress plots show
$\|\boldsymbol{\sigma}\|_F
=(\sum_{i,j}\sigma_{ij}^2)^{1/2}$.
Matrix coloring uses the local matrix Cauchy stress, while beam coloring is a
reconstructed centerline diagnostic rather than a pointwise
three-dimensional fiber stress. Neither quantity should be confused with
the energetic homogenized stress
$\overline{\boldsymbol{\sigma}}_{\mathrm{RVE}}$.

The circular-section integration rule in
Eq.~\eqref{eq:section-quadrature-rule} is characterized in
Table~\ref{tab:section-quadrature-verification}.

\begin{table}[H]
\centering
\caption{Elementary moment audit for the circular-section quadrature rule at
$r=0.05~\mathrm{m}$. Raw area, first moments, and second and cross moments
have units of $\mathrm{m}^2$, $\mathrm{m}^3$, and $\mathrm{m}^4$,
respectively. Errors are normalized by $A_b$, $A_b r$, and
$I=\pi r^4/4$. Area, first moments, and the cross moment are recovered to
roundoff, as are the quadratic moments.}
\label{tab:section-quadrature-verification}
\begin{adjustbox}{max width=\tablewidth,center}
\begin{tabular}{llrrr}
\toprule
Moment integral & units & discrete value & circular value & normalized error \\
\midrule
$1$ & $\mathrm{m}^2$ & $7.853982\times10^{-3}$ & $7.853982\times10^{-3}$ & $-6.626156\times10^{-16}$ \\
$x_n$ & $\mathrm{m}^3$ & $-3.726945\times10^{-20}$ & $0$ & $-9.490587\times10^{-17}$ \\
$x_b$ & $\mathrm{m}^3$ & $-3.388132\times10^{-21}$ & $0$ & $-8.627807\times10^{-18}$ \\
$x_n^2$ & $\mathrm{m}^4$ & $4.908739\times10^{-6}$ & $4.908739\times10^{-6}$ & $3.451123\times10^{-16}$ \\
$x_b^2$ & $\mathrm{m}^4$ & $4.908739\times10^{-6}$ & $4.908739\times10^{-6}$ & $5.176684\times10^{-16}$ \\
$x_nx_b$ & $\mathrm{m}^4$ & $-2.646978\times10^{-23}$ & $0$ & $-5.392379\times10^{-18}$ \\
\bottomrule

\end{tabular}
\end{adjustbox}
\end{table}

Table~\ref{tab:network-metadata} summarizes geometric and discretization
descriptors for the architectures considered below. A physical branch is
distinct from a beam element, which is a numerical subdivision of that
branch. For the Voronoi networks, the branch definition and periodic graph
construction are given in Appendix~\ref{app:voronoi-generation}. To
characterize the range of local geometric scales present in each
architecture, we report the minimum branch-length-to-diameter ratio
$L_{\min}/d$ and the minimum sampled centerline-curvature ratio
$R_{\min}/r$, where $d=2r$. The latter is evaluated from circumradii of
consecutive centerline points away from graph junctions.

The one-dimensional beam approximation is best suited to branches whose
length and radius of curvature are large relative to the fiber diameter.
These geometric ratios cannot be controlled independently in the present
Voronoi construction, and some branches therefore locally violate the
assumptions underlying the beam reduction. The corresponding calculations
are interpreted at the network scale and through homogenized quantities.

\begin{table}[H]
\centering
\caption{Geometric and discretization descriptors for the quantitative
network architectures. Here, $A$ and $\ell$ denote the sinusoidal amplitude
and wavelength, $f_b^{\mathrm{nom}}$ is the nominal geometric fiber fraction
defined in Eq.~\eqref{eq:nominal-fiber-volume-fraction}, $L_{\min}/d$ is the minimum physical-branch
length normalized by the fiber diameter $d=2r$, and $R_{\min}/r$ is the
minimum sampled centerline curvature radius normalized by the fiber radius.
The latter two quantities characterize the local geometric scales represented
by the beam model.
}
\label{tab:network-metadata}
\begin{adjustbox}{max width=\tablewidth,center}
\begin{tabular}{@{}lrrrrrrr@{}}
\toprule
Architecture & $A$ (m) & $\ell$ (m) & branches & elements & $f_b^{\rm nom}$ (\%) & $L_{\min}/d$ & $R_{\min}/r$ \\
\midrule
Single fiber & --- & --- & 1 & 8 & 0.79 & 10.00 & --- \\
Aligned fibers & --- & --- & 30 & 240 & 23.56 & 10.00 & --- \\
Straight cross & --- & --- & 3 & 24 & 2.36 & 10.00 & --- \\
Curved cross & 0.075 & 0.500 & 3 & 120 & 2.81 & 11.93 & 1.74 \\
Wavy aligned & 0.025 & 1.000 & 12 & 120 & 9.48 & 10.06 & 22.13 \\
Isotropic Voronoi & --- & --- & 1342 & 4619 & 4.55 & 1.00 & 1.08 \\
Preferred Voronoi & --- & --- & 702 & 3103 & 2.97 & 1.00 & 1.16 \\
Perforated Voronoi & --- & --- & 1374 & 4766 & 4.61 & 1.02 & 1.16 \\
\bottomrule
\end{tabular}

\end{adjustbox}
\end{table}

\section{Resolved Three-Dimensional Single-Fiber Comparison}
\label{app:single-fiber-resolved-reference}

The mixed-dimensional formulation represents the fiber as an additive
embedded reinforcement: the matrix remains defined over the full domain and
the beam energy is superposed along the fiber centerline. A body-fitted
three-dimensional model instead assigns the fiber its own finite volume and
removes the matrix from that region. The two models therefore do not
represent the same energy functional. We compare them here for the straight
single-fiber geometry of Fig.~\ref{fig:single_fiber_mesh}, where a resolved
fiber remains practical.

Both models are subjected to the constant-volume loading path in
Eq.~\eqref{eq:single-fiber-convergence-load} with
$\lambda_\star=1.10$. The resolved calculation uses a mesh size
$h_{\mathrm{ref}}=0.05~\mathrm{m}$. For the embedded model, the homogenized
affine response under this loading is
\begin{equation}
\overline{\boldsymbol{\sigma}}_{\mathrm{aff}}
=
\mu_m
\operatorname{dev}
\left(
\mathbf F_{\mathrm{macro}}
\mathbf F_{\mathrm{macro}}^T
\right)
+
\frac{A_bL}{|\Omega_0|}
\mu_b
\left(
\lambda^2-\lambda^{-1}
\right)
\mathbf e_1\otimes\mathbf e_1 .
\label{eq:single-fiber-affine-response}
\end{equation}
The embedded calculations reproduce this response to numerical precision
over the sampled matrix and beam discretizations. Let
$\overline{\boldsymbol{\sigma}}_{\mathrm{emb}}$ and
$\overline{\boldsymbol{\sigma}}_{\mathrm{ref}}$ denote the complete
homogenized Cauchy stresses of the embedded and body-fitted models,
respectively. We compare their stress magnitudes and complete tensors using
\begin{equation}
\begin{aligned}
e_n
&=
\frac{
\left|
\|\overline{\boldsymbol{\sigma}}_{\mathrm{emb}}\|_F
-
\|\overline{\boldsymbol{\sigma}}_{\mathrm{ref}}\|_F
\right|
}{
\|\overline{\boldsymbol{\sigma}}_{\mathrm{ref}}\|_F
},
\\
e_T
&=
\frac{
\|
\overline{\boldsymbol{\sigma}}_{\mathrm{emb}}
-
\overline{\boldsymbol{\sigma}}_{\mathrm{ref}}
\|_F
}{
\|\overline{\boldsymbol{\sigma}}_{\mathrm{ref}}\|_F
}.
\end{aligned}
\label{eq:single-fiber-resolved-differences}
\end{equation}
At the loading endpoint, $e_n=2.14\%$ and $e_T=2.75\%$.

\begin{figure}[htb]
\centering
\begin{subfigure}[t]{0.49\linewidth}
\centering
\includegraphics[width=\linewidth]
{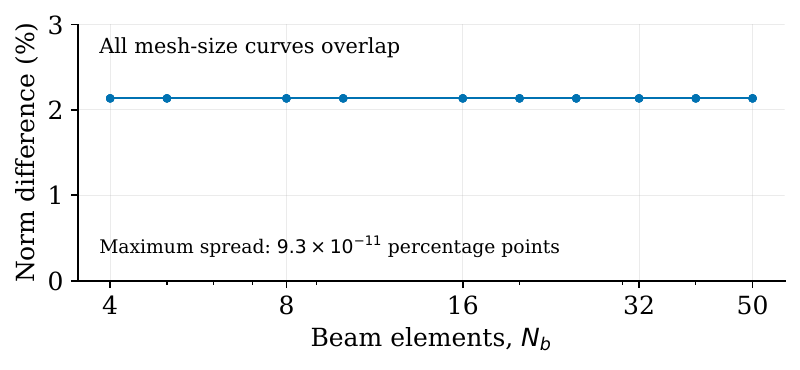}
\caption{Beam element count $N_b$.}
\end{subfigure}
\hfill
\begin{subfigure}[t]{0.49\linewidth}
\centering
\includegraphics[width=\linewidth]
{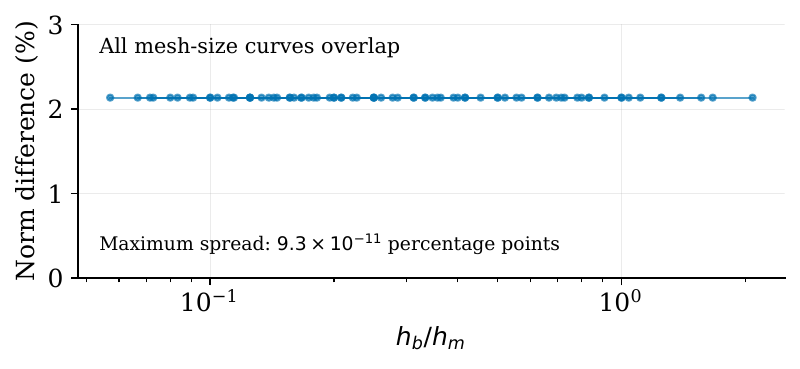}
\caption{Beam-to-matrix resolution ratio
$\rho_h=h_b/h_m$.}
\end{subfigure}
\caption{Stress-norm difference between the mixed-dimensional and
body-fitted single-fiber models over the embedded discretization sweep.
The difference is essentially unchanged as the embedded beam and matrix
discretizations are varied.}
\label{fig:single-fiber-resolved-comparison}
\end{figure}

\begin{table}[H]
\centering
\caption{Summary of the straight-fiber comparison between the
mixed-dimensional and body-fitted three-dimensional models.}
\label{tab:matched-convergence-summary}
\begin{adjustbox}{max width=\tablewidth,center}
\begin{tabular}{@{}ll@{}}
\toprule
Quantity & Value \\
\midrule
Target stretch, $\lambda_\star$ & $1.10$ \\
Resolved mesh size, $h_{\mathrm{ref}}$ & $0.05~\mathrm{m}$ \\
Stress-norm difference, $e_n$ & $2.14\%$ \\
Full-tensor difference, $e_T$ & $2.75\%$ \\
Maximum affine relative tensor error & $4.22\times10^{-10}$ \\
Maximum deformation-component mismatch & $4.41\times10^{-14}$ \\
Representative embedded $(h_m,N_b)$ & $(0.2~\mathrm{m},8)$ \\
\bottomrule
\end{tabular}
\end{adjustbox}
\end{table}

The persistence of the difference over the embedded discretization sweep
indicates that it is not caused by the sampled beam--matrix discretization.
Because the straight-fiber response is affine, the embedded solution is
independent of the section quadrature rule used in this test.

The body-fitted calculation is itself a finite-mesh reference, so the
observed difference reflects both the representation difference and any
remaining discretization error in the resolved model.

\section{Verification and Sensitivity Data}
\label{app:verification-data}

This appendix collects the detailed numerical data supporting the
verification and sensitivity studies in
Sec.~\ref{sec:numerical-verification}. Quantities and comparison measures
retain the definitions introduced in the corresponding main-text studies
unless stated otherwise.

The complete matrix-interpolation and volumetric-treatment study from
Sec.~\ref{sec:higher-order-jbar} is summarized in
Table~\ref{tab:jbar-sensitivity}. The study combines linear (P1) and
quadratic (P2) matrix interpolation, coarse and fine matrix meshes,
$K_m/\mu_m\in\{10,100,1000\}$, and J-bar averaging enabled or disabled.

\begin{table}[H]
\centering
\caption{Factorial coverage and full-tensor stress sensitivities for
the matrix-discretization study. Here, $d_h$ denotes the coarse-to-fine mesh
change and $d_J$ the J-bar on/off difference on the coarse P2 mesh.}
\label{tab:jbar-sensitivity}

\begin{adjustbox}{max width=\tablewidth,center}
\begin{tabular}{@{}ll@{}}
\toprule
Factor & Levels \\
\midrule
Matrix interpolation & P1, P2 \\
Matrix resolution & coarse, fine \\
J-bar averaging & on, off \\
$K_m/\mu_m$ & $10,\ 100,\ 1000$ \\
\bottomrule
\end{tabular}
\end{adjustbox}

\vspace{0.8em}

\begin{adjustbox}{max width=\tablewidth,center}
\begin{tabular}{@{}crrrr@{}}
\toprule
$K_m/\mu_m$
& P1 $d_h$
& P2 $d_h$, on
& P2 $d_h$, off
& P2 $d_J$, coarse \\
\midrule
10   & 5.36\% & 2.48\% & 2.76\% & 0.47\% \\
100  & 5.78\% & 2.61\% & 3.09\% & 0.92\% \\
1000 & 5.48\% & 2.63\% & 3.27\% & 1.25\% \\
\bottomrule
\end{tabular}
\end{adjustbox}

\end{table}
The matrix-mesh resolution study in
Sec.~\ref{sec:matrix-mesh-resolution} compares each mesh with the finest
tested resolution, $h_f=0.10~\mathrm{m}$. In addition to the electrical
measure defined in Eq.~\eqref{eq:matrix-mesh-sensitivity}, the corresponding
stress difference is
\begin{equation}
e_P(h)
=
\frac{
\left\|
\overline{\mathbf P}_{\mathrm{RVE},h}
-
\overline{\mathbf P}_{\mathrm{RVE},h_f}
\right\|_F
}{
\left\|
\overline{\mathbf P}_{\mathrm{RVE},h_f}
\right\|_F
},
\qquad
e_D(h)
=
\frac{
\left\|
\overline{\mathbf D}_{\mathrm{RVE},h}
-
\overline{\mathbf D}_{\mathrm{RVE},h_f}
\right\|_2
}{
\left\|
\overline{\mathbf D}_{\mathrm{RVE},h_f}
\right\|_2
}.
\end{equation}
The differences are evaluated relative to $h_f=0.10~\mathrm{m}$.

\begin{table}[H]
\centering
\caption{Curved-cell response under coupled shear and fixed referential bias
for the matrix-mesh resolution study. Differences are measured relative to
the finest tested matrix mesh $h_f=0.10~\mathrm{m}$, so the zero entries in
that row hold by definition.}
\label{tab:nonaffine-reference-verification}
\begin{adjustbox}{max width=\tablewidth,center}
\begin{tabular}{ccccc}
\toprule
$h$ (m) & $P_{12}$ (MPa) & $D_3$ (C/m$^2$) & $e_P$ & $e_D$ \\
\midrule
$0.20$ & $3.1492910\times10^{-3}$ & $6.3614613\times10^{-4}$ & $0.825\%$ & $0.571\%$ \\
$0.15$ & $3.0973489\times10^{-3}$ & $6.3516535\times10^{-4}$ & $0.319\%$ & $0.415\%$ \\
$0.10$ & $3.0653965\times10^{-3}$ & $6.3254898\times10^{-4}$ & $0.000\%$ & $0.000\%$ \\
\bottomrule

\end{tabular}
\end{adjustbox}
\end{table}

The relative beam--matrix-mesh placement study of
Sec.~\ref{sec:placement-sensitivity} keeps the physical network fixed and translates the
matrix mesh relative to the network while holding all other settings fixed.
The response changes are measured relative to the unshifted configuration.

\begin{table}[H]
\centering
\caption{Relative beam--matrix-mesh placement sensitivity at fixed physical
graph, beam radius, matrix resolution, material parameters, and coupled loading.
The listed offset is the network displacement relative to the mesh, and
$e_P$ and $e_D$ compare
the homogenized stress and electric-displacement responses with the unshifted
configuration.}
\label{tab:placement-sensitivity}
\begin{adjustbox}{max width=\tablewidth,center}
\begin{tabular}{ccccc}
\toprule
Offset (m) & $P_{11}$ (MPa) & $D_3$ (C/m$^2$) & $e_P$ & $e_D$ \\
\midrule
$(0.000,0.000,0.000)$ & $-3.4416331\times10^{-3}$ & $6.9614033\times10^{-4}$ & $0.000\%$ & $0.000\%$ \\
$(0.037,0.061,-0.043)$ & $-3.4418685\times10^{-3}$ & $6.9597101\times10^{-4}$ & $0.047\%$ & $0.103\%$ \\
$(0.091,-0.074,0.058)$ & $-3.4409895\times10^{-3}$ & $6.9555121\times10^{-4}$ & $0.069\%$ & $0.120\%$ \\
\bottomrule

\end{tabular}
\end{adjustbox}
\end{table}

Finally, Table~\ref{tab:sensing-sensitivity} reports the numerical
sensitivity of the fixed-bias observable used for the irregular networks in
Sec.~\ref{sec:voronoi-electrical}. Each value is normalized by the
equilibrated initially biased state of the same calculation, consistent with
Eq.~\eqref{eq:sensor-rve-D-change}. Additional calculations vary either the quadrature rule or the penalty parameter while holding the remaining settings fixed.

\begin{table}[H]
\centering
\caption{ Endpoint normalized fixed-bias sensing signals for the nominally
isotropic, preferred-direction, and perforated networks under the baseline
settings and with independent variations of quadrature and penalty
parameter.}
\label{tab:sensing-sensitivity}
\begin{adjustbox}{max width=\tablewidth,center}
\begin{tabular}{@{}lrrr@{}}
\toprule
Setting & Isotropic & Preferred & Perforated \\
\midrule
Baseline & 9.647778 & 9.752434 & 9.610989 \\
Higher quadrature & 9.647844 & 9.752435 & 9.610989 \\
$p=10^5~\mathrm{MPa}$ & 9.647610 & 9.752571 & 9.614318 \\
$p=10^7~\mathrm{MPa}$ & 9.647878 & 9.752538 & 9.609809 \\
\bottomrule
\end{tabular}

\end{adjustbox}
\end{table}

\section{Voronoi Network Construction}
\label{app:voronoi-generation}

The irregular networks used in Sec.~\ref{sec:irregular-networks} are constructed as periodic Voronoi-based centerline graphs. Random sites are placed in the unit cell using a seeded Mersenne Twister and replicated in a $3\times3\times3$ periodic tiling. A Delaunay triangulation is constructed from the tiled sites, and the corresponding finite Voronoi dual edges are clipped to the central periodic cell. Duplicate segments and segments shorter than $0.02~\mathrm{m}$ are removed before any architecture-specific filters are applied. The target graph connectivity is 5 and the maximum retained connectivity is 6. Geometrically coincident ports are then connected. If several disconnected components remain, they are ranked first by the dimension of the span of their periodic jump vectors and then by vertex count, and the highest-ranked component is retained. At a Voronoi junction, incident branches share translational degrees of freedom but retain independent rotations, giving translationally connected rotational hinges. This differs from the structured center-cross networks, where both translation and rotation are shared at the common junction. A physical branch is defined as a maximal centerline chain between junctions after coincident and opposite-periodic-face endpoints have been identified, while beam elements are numerical subdivisions of these branches.

The three architectures considered in Sec.~\ref{sec:irregular-networks} are obtained by applying different geometric restrictions to this construction. The nominally isotropic network is generated without an imposed orientation filter and therefore represents a finite random realization without a prescribed directional bias. No statistical isotropy is implied. For the preferred-direction network, retained segments with unit tangent $\mathbf t$ satisfy Eq.~\eqref{eq:voronoi-preferred-direction-filter} with $t_z^{\max}=0.7$, which biases the network toward the $x$--$y$ plane. Because this filtering removes segments, it also changes the connectivity, total centerline length, and nominal geometric fiber fraction $f_b^{\mathrm{nom}}$, so the resulting network is not an orientation-only modification at matched fiber content. The perforated network introduces a spherical exclusion region of radius $r_v=0.15~\mathrm{m}$ at the cell center. Candidate centerlines must maintain a clearance of $0.01~\mathrm{m}$ from this region, with beam radius $r=0.01~\mathrm{m}$, so the retained network passes around the central void. The matrix mesh contains the same internal exclusion. Its nominally spherical boundary is represented by the faceted surface of the straight-sided tetrahedral mesh, so uniform refinement preserves the existing boundary geometry rather than converging the spherical representation itself.

\begin{figure}[tbp]
    \centering
    \begin{subfigure}[t]{0.31\linewidth}
        \centering
        \begin{alignedpanel}
        \includegraphics[width=\linewidth,height=0.85\linewidth,keepaspectratio]
        {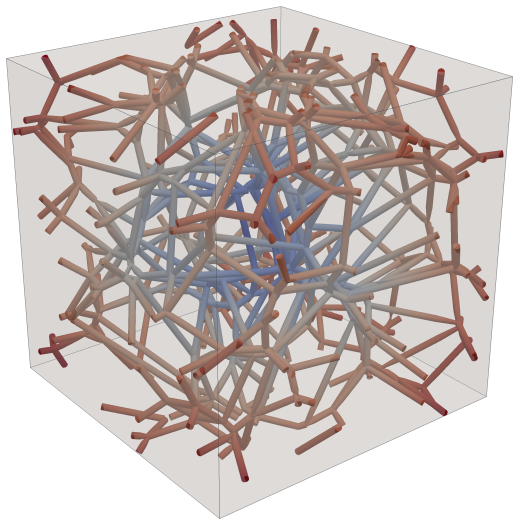}
        \end{alignedpanel}
        \caption{$n=64$, elements $=2612$,\\
        $f_b^{\mathrm{nom}}=2.91\%$.}
    \end{subfigure}\hfill
    \begin{subfigure}[t]{0.31\linewidth}
        \centering
        \begin{alignedpanel}
        \includegraphics[width=\linewidth,height=0.85\linewidth,keepaspectratio]
        {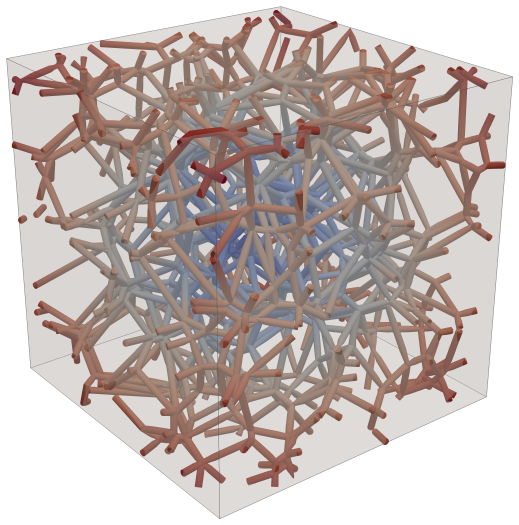}
        \end{alignedpanel}
        \caption{$n=128$, elements $=4627$,\\
        $f_b^{\mathrm{nom}}=4.55\%$.}
    \end{subfigure}\hfill
    \begin{subfigure}[t]{0.31\linewidth}
        \centering
        \begin{alignedpanel}
        \includegraphics[width=\linewidth,height=0.85\linewidth,keepaspectratio]
        {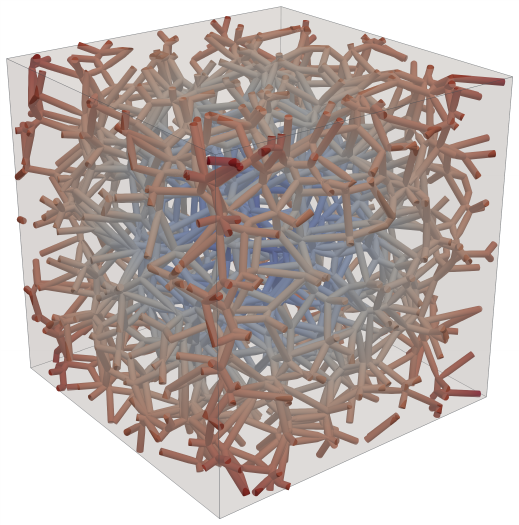}
        \end{alignedpanel}
        \caption{$n=256$, elements $=8086$,\\
        $f_b^{\mathrm{nom}}=7.07\%$.}
    \end{subfigure}

    \vspace{0.6em}

    \begin{subfigure}[t]{0.31\linewidth}
        \centering
        \begin{alignedpanel}
        \includegraphics[width=\linewidth,height=0.85\linewidth,keepaspectratio]
        {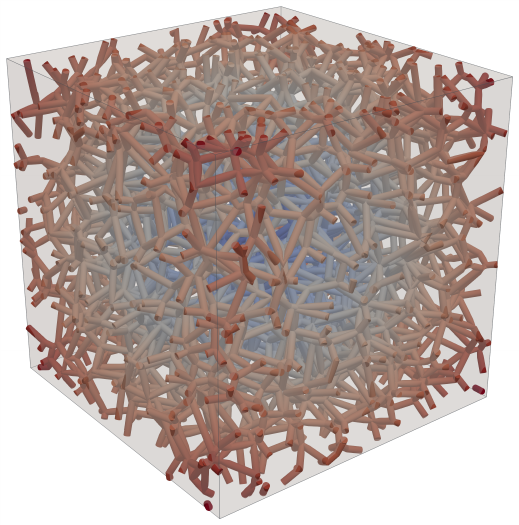}
        \end{alignedpanel}
        \caption{$n=512$, elements $=13591$,\\
        $f_b^{\mathrm{nom}}=10.95\%$.}
    \end{subfigure}\hspace{0.05\linewidth}
    \begin{subfigure}[t]{0.31\linewidth}
        \centering
        \begin{alignedpanel}
        \includegraphics[width=\linewidth,height=0.85\linewidth,keepaspectratio]
        {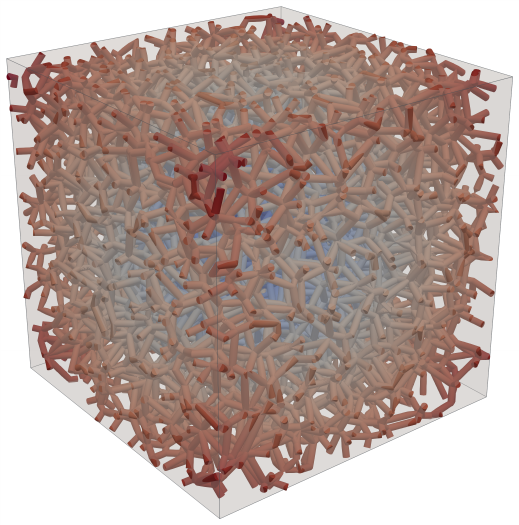}
        \end{alignedpanel}
        \caption{$n=1024$, elements $=21474$,\\
        $f_b^{\mathrm{nom}}=16.59\%$.}
    \end{subfigure}

    \caption{Representative periodic Voronoi networks generated with increasing
    site count at fixed fiber radius and network-generation parameters.}
    \label{fig:voronoi_network_gallery}
\end{figure}

The geometric descriptors reported in Table~\ref{tab:network-metadata} use the same physical-branch definition. The minimum branch slenderness is $L_{\min}/d$, where $d=2r$, and $R_{\min}/r$ uses the smallest circumradius obtained from consecutive centerline points away from graph junctions. 
Figure~\ref{fig:voronoi_network_gallery} shows representative networks generated with increasing site count to illustrate the geometric complexity produced by the construction. 

\clearpage
\section{Homogenized RVE Stress Tensors}
\label{app:rve-stress-tensors}

The complete homogenized RVE Cauchy-stress tensors for the mechanical
loading cases discussed in Sec.~\ref{sec:architecture-response} are reported
below. The displayed tensors are component-wise symmetrized for reporting
and correspond to the final states of the loading paths defined in the main text.

\begin{table}[H]
\centering
\caption{Homogenized RVE Cauchy stresses for the aligned-fiber examples at
the final loading state, in kPa (four significant digits).}
\label{tab:transverse_stress_tensors}
\begin{adjustbox}{max width=\tablewidth,center}
\setlength{\tabcolsep}{6pt}
\renewcommand{\arraystretch}{1.35}
\begin{tabular}{@{}lll@{}}
\toprule
Geometry & Final loading & $\boldsymbol{\sigma}_{\mathrm{RVE}}$ (kPa) \\
\midrule

30 aligned fibers
&
$F_{11}=1.1$
&
$\displaystyle
\begin{bmatrix}
27.72 & 0 & 0 \\
0 & -2.006 & 0 \\
0 & 0 & -2.006
\end{bmatrix}$
\\[1.0em]

30 aligned fibers
&
$F_{22}=1.1$
&
$\displaystyle
\begin{bmatrix}
-13.02 & 0 & 0 \\
0 & 4.012 & 0 \\
0 & 0 & -2.006
\end{bmatrix}$
\\[1.0em]

30 aligned fibers
&
$F_{12}=0.25$
&
$\displaystyle
\begin{bmatrix}
0.8333 & 5.000 & 0 \\
5.000 & -0.4167 & 0 \\
0 & 0 & -0.4167
\end{bmatrix}$
\\

\bottomrule
\end{tabular}
\end{adjustbox}
\end{table}

\begin{table}[H]
\centering
\caption{Homogenized RVE Cauchy stresses for the lattice and wavy-fiber
examples at the final loading state, in kPa (four significant digits).}
\label{tab:lattice_stress_tensors}
\begin{adjustbox}{max width=\tablewidth,center}
\setlength{\tabcolsep}{6pt}
\renewcommand{\arraystretch}{1.35}
\begin{tabular}{@{}lll@{}}
\toprule
Geometry & Final loading & $\boldsymbol{\sigma}_{\mathrm{RVE}}$ (kPa) \\
\midrule

Straight cross
&
$F_{11}=1.1$
&
$\displaystyle
\begin{bmatrix}
4.803 & 0 & 0 \\
0 & -2.373 & 0 \\
0 & 0 & -2.373
\end{bmatrix}$
\\[1.0em]

Curved cross
&
$F_{11}=1.1$
&
$\displaystyle
\begin{bmatrix}
4.358 & 0.009211 & -4.804\times10^{-4} \\
0.009211 & -2.094 & 0.004089 \\
-4.804\times10^{-4} & 0.004089 & -2.173
\end{bmatrix}$
\\[1.0em]

Wavy aligned
&
$F_{11}=1.1$
&
$\displaystyle
\begin{bmatrix}
13.14 & 4.116\times10^{-6} & -2.984\times10^{-5} \\
4.116\times10^{-6} & -1.961 & -1.513\times10^{-5} \\
-2.984\times10^{-5} & -1.513\times10^{-5} & -2.010
\end{bmatrix}$
\\[1.0em]

Straight cross
&
$F_{12}=0.25$
&
$\displaystyle
\begin{bmatrix}
0.8533 & 5.083 & -8.396\times10^{-4} \\
5.083 & -0.1958 & -5.495\times10^{-4} \\
-8.396\times10^{-4} & -5.495\times10^{-4} & -0.4167
\end{bmatrix}$
\\[1.0em]

Curved cross
&
$F_{12}=0.25$
&
$\displaystyle
\begin{bmatrix}
0.9192 & 5.254 & 0.002467 \\
5.254 & -0.4175 & -0.001617 \\
0.002467 & -0.001617 & -0.3977
\end{bmatrix}$
\\[1.0em]

Wavy aligned
&
$F_{12}=0.25$
&
$\displaystyle
\begin{bmatrix}
0.7611 & 5.093 & -3.491\times10^{-5} \\
5.093 & -0.4222 & 7.994\times10^{-5} \\
-3.491\times10^{-5} & 7.994\times10^{-5} & -0.4198
\end{bmatrix}$
\\

\bottomrule
\end{tabular}
\end{adjustbox}
\end{table}

\begin{table}[H]
\centering
\caption{Homogenized RVE Cauchy stresses for the Voronoi examples at the
final loading state, in kPa (four significant digits).}
\label{tab:voronoi_stress_tensors}
\begin{adjustbox}{max width=\tablewidth,center}
\setlength{\tabcolsep}{6pt}
\renewcommand{\arraystretch}{1.35}
\begin{tabular}{@{}lll@{}}
\toprule
Geometry & Final loading & $\boldsymbol{\sigma}_{\mathrm{RVE}}$ (kPa) \\
\midrule

Isotropic
&
$F_{11}=1.1$
&
$\displaystyle
\begin{bmatrix}
4.698 & -0.02229 & -0.006573 \\
-0.02229 & -2.271 & -0.007990 \\
-0.006573 & -0.007990 & -2.230
\end{bmatrix}$
\\[1.0em]

Isotropic
&
$F_{33}=1.1$
&
$\displaystyle
\begin{bmatrix}
-2.234 & 0.01669 & 0.01282 \\
0.01669 & -2.257 & -0.01076 \\
0.01282 & -0.01076 & 4.678
\end{bmatrix}$
\\[1.0em]

Isotropic
&
$F_{12}=0.1$
&
$\displaystyle
\begin{bmatrix}
0.1518 & 2.291 & -0.008741 \\
2.291 & -0.05236 & 0.007757 \\
-0.008741 & 0.007757 & -0.05694
\end{bmatrix}$
\\[1.0em]

Preferred
&
$F_{11}=1.1$
&
$\displaystyle
\begin{bmatrix}
4.699 & 0.007410 & -0.01250 \\
0.007410 & -2.233 & -0.01588 \\
-0.01250 & -0.01588 & -1.989
\end{bmatrix}$
\\[1.0em]

Preferred
&
$F_{22}=1.1$
&
$\displaystyle
\begin{bmatrix}
-2.229 & -0.01131 & 0.006094 \\
-0.01131 & 4.702 & 0.03480 \\
0.006094 & 0.03480 & -1.989
\end{bmatrix}$
\\[1.0em]

Preferred
&
$F_{33}=1.1$
&
$\displaystyle
\begin{bmatrix}
-2.316 & 0.002795 & 0.002744 \\
0.002795 & -2.315 & -0.01738 \\
0.002744 & -0.01738 & 3.999
\end{bmatrix}$
\\[1.0em]

Preferred
&
$F_{12}=0.1$
&
$\displaystyle
\begin{bmatrix}
0.1678 & 2.235 & 0.002761 \\
2.235 & -0.06136 & -0.006570 \\
0.002761 & -0.006570 & -0.06699
\end{bmatrix}$
\\[1.0em]

Perforated
&
$F_{11}=1.1$
&
$\displaystyle
\begin{bmatrix}
4.718 & 0.01366 & 0.03242 \\
0.01366 & -2.212 & -0.008230 \\
0.03242 & -0.008230 & -2.201
\end{bmatrix}$
\\[1.0em]

Perforated
&
$F_{22}=1.1$
&
$\displaystyle
\begin{bmatrix}
-2.263 & -0.006086 & -0.01152 \\
-0.006086 & 4.680 & 0.02431 \\
-0.01152 & 0.02431 & -2.262
\end{bmatrix}$
\\[1.0em]

Perforated
&
$F_{33}=1.1$
&
$\displaystyle
\begin{bmatrix}
-2.253 & -0.006088 & -0.01736 \\
-0.006088 & -2.266 & -0.01116 \\
-0.01736 & -0.01116 & 4.668
\end{bmatrix}$
\\[1.0em]
    
    Perforated
    &
    $F_{13}=0.1$
    &
    $\displaystyle
    \begin{bmatrix}
0.2172 & 0.01157 & 2.285 \\
0.01157 & -0.05002 & 0.002617 \\
2.285 & 0.002617 & -0.04562
\end{bmatrix}$
    \\
    
    \bottomrule
    \end{tabular}
    \end{adjustbox}
    \end{table}

\end{document}